# RHDLPP: A multigroup radiation hydrodynamics code for laser-produced plasmas

Qi Min[1,2,*], Ziyang Xu[1], Siqi He[1], Haidong Lu[1], Xingbang Liu[1], Ruizi Shen[1], Yanhong Wu[1], Qikun Pan[3], Chongxiao Zhao[3], Fei Chen[3], Maogen Su[1,2,*], and Chenzhong Dong[1,2,*]

[1]*Key Laboratory of Atomic and Molecular Physics & Functional Material of Gansu Province, College of Physics and Electronic Engineering, Northwest Normal University, Lanzhou 730070, China*
[2]*Gansu International Scientific and Technological Cooperation Base of Laser Plasma Spectroscopy, Lanzhou, 730070, China*
[3]*Changchun Institute of Optics, Fine Mechanics and Physics, Chinese Academy of Sciences, Changchun, 130033, China*

E-mail: mq_lpps@nwnu.edu.cn; sumg@nwnu.edu.cn; dongcz@nwnu.edu.cn

**Abstract**: In this paper, we introduce the RHDLPP, a flux-limited multigroup radiation hydrodynamics numerical code designed for simulating laser-produced plasmas in diverse environments. The code bifurcates into two packages: RHDLPP-LTP for low-temperature plasmas generated by moderate-intensity nanosecond lasers, and RHDLPP-HTP for high-temperature, high-density plasmas formed by high-intensity laser pulses. The core radiation hydrodynamic equations are resolved in the Eulerian frame, employing an operator-split method. This method decomposes the solution into two substeps: first, the explicit resolution of the hyperbolic subsystems integrating radiation and fluid dynamics; second, the implicit treatment of the parabolic part comprising stiff radiation diffusion, heat conduction, and energy exchange. Laser propagation and energy deposition are modeled through a hybrid approach, combining geometrical-optics ray-tracing in sub-critical plasma regions with a one-dimensional solution of the Helmholtz wave equation in super-critical areas. The thermodynamic states are ascertained using an equation of state, based on either the real gas approximation or the quotidian equation of state (QEOS). For ionization calculations, the code employs a steady-state collisional-radiation (CR) model using the screened-hydrogenic approximation. Additionally, RHDLPP includes RHDLPP-SpeIma3D, a three-dimensional spectral simulation post-processing module, for generating both temporally-spatially resolved and time-integrated spectra and imaging, facilitating direct comparisons with experimental data. The paper showcases a series of verification tests to establish the code's accuracy and efficiency, followed by application cases, including simulations of laser-produced aluminium (Al) plasmas, pre-pulse-induced target deformation of tin (Sn) microdroplets relevant to extreme ultraviolet lithography light sources, and varied imaging and spectroscopic simulations. These simulations highlight RHDLPP's effectiveness and applicability in fields such as laser-induced breakdown spectroscopy, extreme ultraviolet lithography sources, and high-energy-density physics.

# 1. Introduction

Upon focusing a high-energy laser pulse onto the surface of a target material, typically solid or liquid, a laser-produced plasma (LPP) forms, characterized by its high temperature and density. The LPP, a potent source of ions, electrons, and both neutral and excited particles, emits intense radiation across the visible, ultraviolet (UV), extreme ultraviolet (EUV), and X-ray spectrums. As a result, it finds extensive applications in various fields, including pulsed-laser thin-film deposition [1], laser-induced breakdown spectroscopy (LIBS) [2, 3], astrophysical plasma simulations [4], inertial confinement fusion (ICF) [5], extreme ultraviolet lithography (EUVL) sources [6], laboratory ion sources [7], and high-energy-density physics (HEDP) [8].

The LPP exhibits high temperatures and high electron and ion densities within compact dimensions, alongside spatial non-uniformity, rapid time-dependent changes, broad-spectrum electromagnetic radiation, and significant opacity effects on radiation [9]. Its formation and evolution encompass a complex interplay and competition among numerous physical processes [10], such as the interaction between macroscopic dynamic evolution and microscopic atomic processes, and the energy exchange between radiation and matter. Given these complexities, employing a single experimental method for a comprehensive multi-parameter state diagnosis of LPP is often unrealistic [11]. Moreover, the intricate plasma properties pose challenges in interpreting and understanding experimental data; for instance, non-uniformities and opacities significantly interfere with typical spectroscopic measurements [11]. Numerical simulation codes based on radiation hydrodynamics (RHD) models serve as a powerful complement to experimental measurements, aiding not only in the analysis and simulation of experimental results but also in generating experimentally testable predictions.

A comprehensive and self-consistent RHD code for LPP simulation necessitates the inclusion of the dynamic evolution of the target and plasma, laser energy deposition, radiative transfer, interactions between radiation and matter, heat conduction, and the interactions among various microscopic particles (e.g., electrons, ions, photons) within the plasma. Over the past decades, primarily driven by the demands in the fields of ICF, EUVL sources, and HEDP, numerous RHD codes for LPPs have emerged. These codes, in terms of primary methods for numerically solving RHD equations, are categorized into three types: Eulerian, Lagrangian, and Arbitrary Lagrangian-Eulerian (ALE) codes.

In Eulerian-type codes, the computational mesh remains fixed while the continuum moves relative to the grid. This setup facilitates the handling of large distortions in continuum motion with relative ease, albeit often at the expense of precise interface definition and resolution of flow details. Representative codes employing the Eulerian approach for LPP simulation include FLASH [12], CRASH [13], and xRAGE [14, 15]. These codes treat the plasma as a single fluid with three temperatures (electron temperature $T_e$, ion temperature $T_i$, and radiation temperature $T_r$), incorporating adaptive mesh refinement (AMR) for efficient simulations. Conversely, Lagrangian methods, wherein each node of the computational mesh follows the associated material particle during motion, facilitate easy tracking of free surfaces and material interfaces. This method also accommodates the treatment of materials with history-dependent constitutive relations. However, it struggles with large distortions of the computational domain without frequent remeshing operations. Codes employing the Lagrangian approach for LPP simulation encompass the one-dimensional (1D), three fluid code HYADES [16], the 1D hydrodynamics code MEDUSA [17], the 1D implicit Lagrangian codes MULTI-1D [18] and MULTI-fs [19] in the MULTI family, the single-fluid, two-temperature two-dimensional (2D) hydrodynamics code ATLANT [20], the 1D radiation-magnetohydrodynamics code HELIOS-CR [21], and LASNEX code[22]. In the ALE description, the nodes of the computational mesh may move with the continuum in a conventional Lagrangian manner, remain fixed in an Eulerian manner, or move in an arbitrarily defined way to provide continuous rezoning capability. This flexibility in mesh movement, offered by the ALE description, allows for handling greater distortions of the continuum compared to a purely Lagrangian method, with better resolution than a purely Eulerian approach. LPP simulation codes based on the ALE method include LARED-Integration [23], HYDRA [24], ICF3D [25], MULTI-2D [26], RALEF-2D [27], PALE [28], CHIC [29], and TRHD [30], among others.

In addition to simulating plasma formation and evolution, and outputting parameters such as temperature and electron density, it is highly beneficial for an RHD code to generate simulations that can be directly compared to experimental measurements. This capability not only aids in benchmarking the code but also facilitates the interpretation of experimental results and the understanding of underlying physical mechanisms. However, experimental measurements often reflect a combination of plasma state parameters. For instance, the intensity and profile of the experimental spectrum correlate with temperature, electron density, and charge state distribution [11]. Therefore, to enable direct comparison with experimental results, RHD codes typically necessitate post-processing capabilities. Several codes already incorporate corresponding post-processing diagnostic modules. For instance, the FLASH code [12] includes diagnostic modules such as proton imaging, proton emission, Thomson scattering, and X-ray imaging. Similarly, the HELIOS code features a post-processing package, SPECT3D [31], capable of generating simulated spectra and images. Additionally, the Finite-Element Spectral Transfer of Radiation (FESTR) code [32] can be employed for post-processing RHD simulations and analyzing experimental spectral data. These advancements in RHD codes significantly enhance their utility in plasma research.

In this paper, we introduce the development of a RHD code for LPPs based on the Eulerian method, referred to as RHDLPP. This code is capable of simulating not only lower-temperature plasmas produced by moderate-intensity ns-laser pulses but also higher-temperature and density LPPs. The code encompasses a range of physical processes, including laser energy deposition, dynamic plasma evolution, radiative transfer, and the interaction between radiation and matter, as well as electron heat conduction. Additionally, the code features a specialized three-dimensional (3D) spectral simulation post-processing module, named RHDLPP-SpeIma3D, which offers temporal-spatially resolved as well as time-integrated simulated spectra and imaging. This ensures that the simulation outcomes can be directly compared with experimental spectral results.

The paper is structured to first introduce the framework of RHDLPP and its solution modules (Section 2), then detail the RHD model, governing equations, and numerical algorithms (Section 3). We present simulation results for classic RHD problems and code verification in Section 4. Section 5 is divided into three subsections: Subsection 5.1 shows the simulation results of aluminum (Al) plasma produced by ns pulses of different intensities and compares with other codes, Subsection 5.2 presents deformation results of liquid droplets from laser impacts on tin (Sn) microdroplets, and Subsection 5.3 discusses imaging and spectral simulation results. The paper concludes in Section 6 with a summary, discussing the code's advantages, limitations, and future improvement plans.

## 2. Code description

RHDLPP is a specialized RHD code, uniquely tailored for LPPs. This single-fluid, two-temperature RHD code operates under the assumption that electron and ion temperatures ($T_e$ and $T_i$, respectively) are equivalent, yet distinct from the radiation temperature ($T_r$). The RHD equations are resolved within an Eulerian framework, accommodating 1D or 2D Cartesian, spherical, or cylindrical coordinates. RHDLPP integrates a suite of modules encompassing hydrodynamics, heat conduction, radiation transport, laser energy deposition, equations of state (EOS), charge state distribution, and spectral simulation. The code adopts the Flux-Limited Diffusion (FLD) approximation [33] for modeling radiation transport, thereby approximating the free-streaming limit.

Laser propagation and energy deposition in RHDLPP are simulated through a hybrid methodology [34] that merges geometrical-optics ray-tracing in sub-critical plasma regions with a 1D solution of the Helmholtz wave equation in super-critical regions. Thermodynamic properties within the code are ascertained using an EOS, based either on the real gas approximation [35] or a quotidian equation of state (QEOS) [36]. The charge state distribution and average ionization degree are calculated employing a steady-state collisional-radiation (CR) model [37]. This model includes various atomic processes derived from the screened-hydrogenic approximation [38]. The RHDLPP-SpeIma3D, a post-processing spectral simulation module, generates images and spectra by solving the radiative transfer equation across multiple lines-of-sight through the plasma.

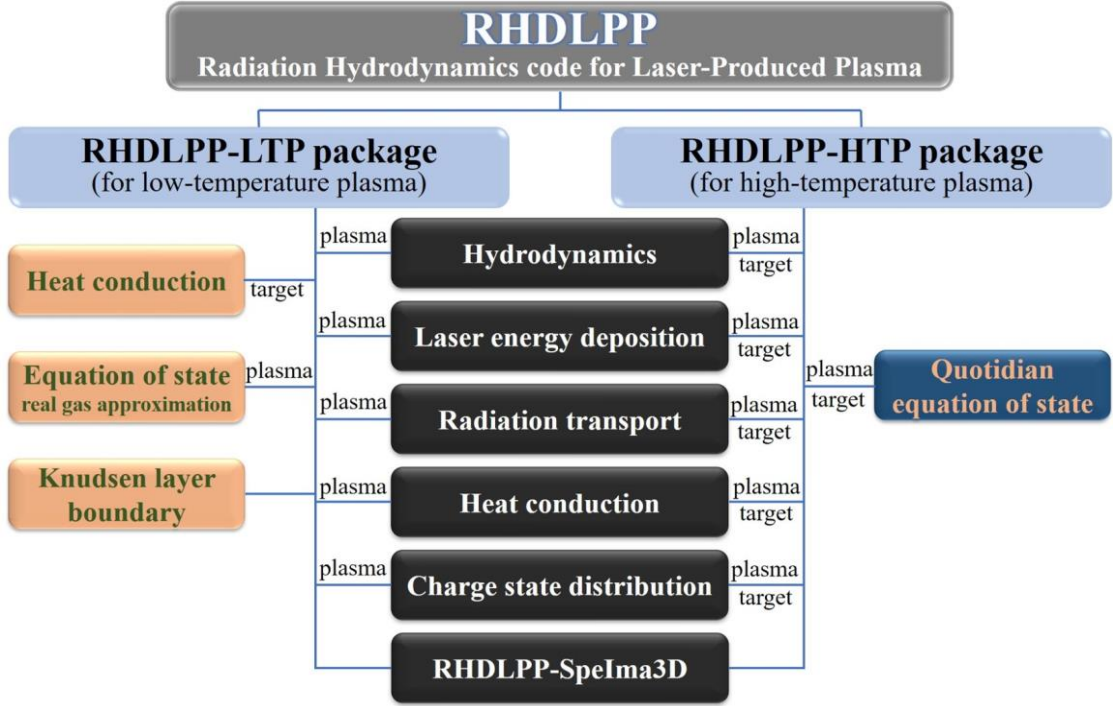

**Fig. 1.** Overall framework and main modules of the RHDLPP code.

The primary framework and modules of RHDLPP are depicted in Fig. 1. The radiation and dynamic characteristics of LPPs, and the physical processes underpinning their formation and evolution, are heavily influenced by the parameters of the ablating laser. Consequently, RHDLPP is bifurcated into two distinct packages: RHDLPP-LTP and RHDLPP-HTP, catering to different laser parameter scenarios.

The RHDLPP-LTP package is designed for simulating low-temperature plasmas produced by moderate-intensity ns lasers. When such a laser, with a power density higher than the target ablation threshold yet below $10^9$ W/cm$^2$ (a rough estimate), irradiates the target, the target heating, melting and evaporation dominated by heat conduction will firstly occur. The vapor produced by the laser pulse front is similar to a transparent thin medium and the laser beam can pass through it to the target surface almost without attenuation. Subsequently, the incident laser radiation is absorbed in the form of inverse Bremsstrahlung by the gradually heated vapor, resulting in vapor breakdown and plasma formation [39]. It is worth noting that in the process of target evaporation caused by such moderate-intensity laser radiation, the target surface temperature generally will not exceed the critical value, defined as the end point of a phase equilibrium curve. Accordingly, a sharp phase boundary will be formed between the solid and vapor phases. This boundary is a transient nonequilibrium thin layer with a thickness of only a few molecular mean free paths, which is called the Knudsen layer (KL) [40]. The state parameters (such as density, temperature, and pressure) of the target surface and the vapor plasma adjacent to the KL may exhibit a discontinuity. The plasma flow above KL can be regarded as a hydrodynamic flow satisfying the continuum approximation. The maximum temperature of the plasma typically remains below 10 eV, while the electron density does not exceed $10^{20}$ cm$^{-3}$. The emitted radiation spectra are mainly in the visible and infrared range. The RHDLPP-LTP package thus segregates the computational region for target material and plasma. Laser energy deposition on the target surface and energy transfer within the target are modeled by a heat conduction equation across the target's computational region. The interaction between the laser and plasma and the plasma's dynamic evolution are addressed by RHD equations within the plasma computational region. The KL acts as a boundary, establishing conditions that couple the heat conduction equation for the target with RHD equations for the vapor plasma and ambient gas phase. The internal energy and pressure of the plasma are defined by an EOS based on the real gas approximation [35].

In scenarios where the laser intensity is sufficiently high, the target temperature can be greater than the critical temperature, so that the sharp interface between the target surface and vapor plasma phases disappears and is smeared into a macroscopic transition layer [41]. Under such conditions, the target material should contribute mass to the plasma region mainly through hydrodynamic expansion, and the laser ablation process should be described by solving the RHD equations for the whole physical domain, supplemented with wide range EOS. Thus, in the RHDLPP-HTP package, which is tailored for simulating high-temperature, high-density plasmas produced by high-intensity laser pulses, the expansion of both target material and plasma is described uniformly by the RHD equations in an overall computational region. Thermodynamic properties are determined using QEOS [36].

As illustrated in Fig. 1, operating the RHDLPP-LTP package involves invoking modules for heat conduction in the target, hydrodynamics, laser energy deposition, radiation transport, real gas approximation EOS, and charge state distribution for the plasma, along with a Knudsen layer boundary module for establishing target-plasma boundary conditions. Conversely, the RHDLPP-HTP package employs modules for hydrodynamics, laser energy deposition, radiation transport, heat conduction, QEOS, and charge state distribution for both the target and plasma. If spectroscopic simulations are required, both packages engage the RHDLPP-SpeIma3D module, using plasma state parameters like temperature and density as inputs.

## 3. Radiation hydrodynamics model and numerical methods

3.1. Multigroup radiation hydrodynamics equations in the flux-limited diffusion limit

In the RHDLPP code, a single-fluid, two-temperature model characterizes the fluids, ensuring that all atomic and ionic species, along with electrons, possess the same macroscopic fluid velocity. The multigroup RHD equations under the FLD approximation [33] solved by RHDLPP to depict the evolution of a two-temperature plasma are provided below [33, 42, 43].

$$\frac{\partial \rho}{\partial t} + \nabla \cdot (\rho \boldsymbol{u}) = 0, \tag{1}$$

$$\frac{\partial}{\partial t}(\rho \boldsymbol{u}) + \nabla \cdot (\rho \boldsymbol{u}\boldsymbol{u}) + \nabla p + \sum_{g=1}^{G} \lambda_g \nabla E_g = 0, \tag{2}$$

$$\frac{\partial}{\partial t}(\rho E) + \nabla \cdot [(\rho E + p)\boldsymbol{u}] + \boldsymbol{u} \cdot \sum_{g=1}^{G} \lambda_g \nabla E_g = Q_L + \nabla \cdot C_e \nabla T + \sum_{g=1}^{G} c(\kappa_{Pa,g} E_g - \kappa_{Pe,g} B_g), \tag{3}$$

$$\frac{\partial E_g}{\partial t} + \nabla \cdot \left(\frac{3-f_g}{2} E_g \boldsymbol{u}\right) - \boldsymbol{u} \cdot \lambda_g \nabla E_g = \nabla \cdot (-\boldsymbol{F}_g) + (c\kappa_{Pe,g} B_g - c\kappa_{Pa,g} E_g). \tag{4}$$

Here, $\rho$, $\boldsymbol{u}$, $T$, and $p$ represent the matter density, velocity, temperature, and pressure of the material, respectively. $E$ is the total matter energy per unit mass (internal energy plus kinetic energy). The electron thermal conductivity is denoted by $C_e$, while $Q_L$ represents the energy source due to laser heating. The radiation energy density in each group and the group energy density in the Planck spectrum are represented by $E_g$ and $B_g$, respectively. $B_g = (15aT^4/\pi^4)[P(x_{g+1}) - P(x_g)]$, where $a$ is the radiation constant, $P(x)$ is the Planck integral, defined as $P(x) = \int_0^x (dx')(x')^3/[\exp(x')-1]$. $\kappa_{R,g}$, $\kappa_{Pe,g}$, and $\kappa_{Pa,g}$ denote the group Rosseland mean coefficient, Planck mean emission coefficient, and Planck mean absorption coefficient, respectively. $\lambda_g$ and $f_g$ are the flux limiter and Eddington factor, respectively. The speed of light is represented by $c$.

In the FLD approximation, the radiation flux $\boldsymbol{F}_g$ in Eq. (4) is written in the form of Fick's law of diffusion [33]

$$\boldsymbol{F}_g = -D_g \nabla E_g, \qquad D_g = \frac{c\lambda_g}{\kappa_{R,g}}. \tag{5}$$

where $D_g$ is the diffusion coefficient. The magnitude of the radiation flux must be at most $cE_g$ to recover the free-streaming limit in the optically thin limit. Various flux limiters have been proposed in the literature [44-46] that ensure the diffusion flux is constrained by this free-streaming flux. We adopt the flux limiter approximation given in Levermore & Pomraning [45] as

$$\lambda_g = \frac{2+R}{6+3R+R^2}, \qquad R = \frac{|\nabla E_g|}{\kappa_{R,g} E_g}, \qquad f_g = \lambda_g + \lambda_g^2 R^2. \tag{6}$$

In the optically thick limit, both the flux limiter $\lambda_g$ and the Eddington factor $f_g$ approach 1/3. On the other hand, in the optically thin limit, the flux limiter $\lambda_g$ and the Eddington factor $f_g$ approach 0 and 1, respectively.

Similarly, we implemented the flux limiter to constrain the electron heat flux [8]. In LPPs, the temperature scale length $L_T = T_e/|\nabla T_e|$ can be smaller than the collisional mean free path of electrons. In such cases, the classical Spitzer-Harm (SH) description for collisional electron conductivity becomes inadequate, and the observed electron heat flux is significantly smaller than the SH value [8]. To address this issue, the electron heat flux $\boldsymbol{F}_e = -C_e \nabla T$ in Eq. (4) is typically limited to a fraction of the free-streaming value $F_{FS}$ by $\boldsymbol{F}_e = -(fF_{FS}/|\nabla T|)\nabla T$, where $F_{FS} = n_e kT\sqrt{(kT/m_e)}$ and $f$ represents an adjustable flux limiter. Consequently, the flux-limited electron heat flux can be expressed as [13]:

$$\boldsymbol{F}_e = -\min\left(C_e, \frac{fF_{FS}}{|\nabla T|}\right)\nabla T. \tag{7}$$

3.2. Numerical method

The governing equations (1-4), under the Flux-Limited Diffusion (FLD) assumption, constitute a mixed hyperbolic-parabolic system with stiff source terms [42, 43]. Our approach employs a commonly used operator splitting scheme, wherein the hyperbolic subsystem is initially solved explicitly, followed by an implicit treatment of the parabolic component, which includes diffusion and source-sink terms.

The hyperbolic subsystem, derived by omitting the right-hand side terms of Eqs. (1)-(4), comprises:

$$\frac{\partial \rho}{\partial t} + \nabla \cdot (\rho \boldsymbol{u}) = 0, \tag{8}$$

$$\frac{\partial}{\partial t}(\rho \boldsymbol{u}) + \nabla \cdot (\rho \boldsymbol{u}\boldsymbol{u}) + \nabla p + \sum_{g=1}^{G} \lambda_g \nabla E_g = 0, \tag{9}$$

$$\frac{\partial}{\partial t}(\rho E) + \nabla \cdot [(\rho E + p)\boldsymbol{u}] + \boldsymbol{u} \cdot \sum_{g=1}^{G} \lambda_g \nabla E_g = 0, \tag{10}$$

$$\frac{\partial E_g}{\partial t} + \nabla \cdot \left(\frac{3-f_g}{2} E_g \boldsymbol{u}\right) - \boldsymbol{u} \cdot \lambda_g \nabla E_g = 0. \tag{11}$$

These partial differential equations of the hyperbolic subsystem can be generalized as:

$$\frac{\partial \boldsymbol{U}}{\partial t} + \nabla \cdot \boldsymbol{F} = \boldsymbol{S}, \tag{12}$$

Here

$$\boldsymbol{U} = \begin{pmatrix} \rho \\ \rho \boldsymbol{u} \\ \rho E \\ E_g \end{pmatrix}, \quad \boldsymbol{F} = \begin{pmatrix} \rho \boldsymbol{u} \\ \rho \boldsymbol{u}\boldsymbol{u} + p \\ \rho E \boldsymbol{u} + p\boldsymbol{u} \\ (3-f_g)E_g \boldsymbol{u}/2 \end{pmatrix}, \quad \boldsymbol{S} = \begin{pmatrix} 0 \\ -\sum_{g=1}^{G} \lambda_g \nabla E_g \\ -\boldsymbol{u} \cdot \sum_{g=1}^{G} \lambda_g \nabla E_g \\ \boldsymbol{u} \cdot \lambda_g \nabla E_g \end{pmatrix}. \tag{13}$$

where $\boldsymbol{U}$ represents the conserved variables, $\boldsymbol{F}$ their flux, and $\boldsymbol{S}$ encompasses the radiation force, work done by the radiation field, and possible geometrical source terms.

For computational efficiency, we utilize Strang operator splitting to alternately treat the source term vector of the hyperbolic subsystem and the spatial differential terms of Eq. (12), achieving a second-order method. Solution operators $\mathcal{L}^{\Delta t}$ and $\mathcal{S}^{\Delta t}$, corresponding to the ideal hydrodynamic equations augmented with the advection of the radiation energy (i.e., $\boldsymbol{S} \equiv \boldsymbol{0}$ in Eq. (12)) and the ordinary differential equation respectively, are used. The solution vector $\boldsymbol{U}^{n+1,*}$ is computed from $\boldsymbol{U}^n$ as follows:

$$\boldsymbol{U}^{n+1,*} = \mathcal{S}^{\Delta t/2} \mathcal{L}^{\Delta t} \mathcal{S}^{\Delta t/2}(\boldsymbol{U}^n). \tag{14}$$

where the $*$ superscript denotes the state following the explicit update.

The hydrodynamic equations are solved using the second-order Total Variation Diminishing (TVD) MUSCL-Hancock scheme [47] (MUSCL stands for Monotonic Upstream-Centred Scheme for Conservation Laws). The conserved variables are defined at cell centers. For simplicity, we assume a 1D Cartesian grid with spacing $\Delta x$, cell center index $i$, and cell face between cells $i$ and $i + 1$ identified by half indices $i + 1/2$.

The TVD MUSCL-Hancock scheme is implemented in the following steps:

Step 1: **Data reconstruction**. The conservative variables at the left and right boundaries of the cell are obtained by the linear boundary extrapolation:

$$U_i^L = U_i - \frac{1}{2}\phi(r)\Delta_i, \qquad U_i^R = U_i + \frac{1}{2}\phi(r)\Delta_i. \tag{15}$$

Here

$$\Delta_i = \Delta U_{i+1/2} + \Delta U_{i-1/2}, \qquad \Delta U_{i+1/2} = U_{i+1} - U_i, \qquad \Delta U_{i-1/2} = U_i - U_{i-1}. \tag{16}$$

where $\phi(r)$ is the slope limiter, $r = \Delta U_{i-1/2}/\Delta U_{i+1/2}$ is the ratio of slopes on each side of the cell. In RHDLPP, the available slope limiters include Minmod, van Leer, van Albada, Superbee, and Monotonized Central (MC) [47].

Step 2: **Evolution of extrapolated values**. The left and right interpolated values from Eq. (15) are then evolved over a half time step:

$$U_i^{n+1/2,L} = U_i^L + \frac{\Delta t}{2}\left[\frac{F(U_i^L) - F(U_i^R)}{\Delta x}\right], \tag{17}$$

$$U_i^{n+1/2,R} = U_i^R + \frac{\Delta t}{2}\left[\frac{F(U_i^L) - F(U_i^R)}{\Delta x}\right], \tag{18}$$

Step 3: **Solution of Riemann problem**. The half-step values are then used in conjunction with a Riemann solver to compute the intercell flux at $t^{n+1/2}$, $F_{i+1/2}^{n+1/2}$. The RHDLPP code currently includes several Riemann solvers for computing fluxes as a function of the reconstructed left and right states: the HLL, HLLE, HLLC [47] solvers, the Roe solver [48], the AUSM+ scheme [49], and the local Lax-Friedrichs solver [50].

Step 4: **Updating of solution**. Once the fluxes have been computed, they are utilized to calculate the ultimate MUSCL-Hancock resolution, which is only an intermediate result.

$$U_i^* = U_i^n + \Delta t\left[\frac{F_{i-1/2}^{n+1/2} - F_{i+1/2}^{n+1/2}}{\Delta x}\right], \tag{19}$$

To ensure the stability of the explicit step, the Courant-Friedrich-Levy (CFL) stability condition used to estimate the timestep $\Delta t$ also takes into account the radiative pressure. The updated CFL condition is simply [43]

$$\Delta t \leq C_{CFL}\frac{\Delta x}{u + \sqrt{c_m^2 + \sum_{g=1}^{G} c_g^2}}. \tag{20}$$

where $C_{CFL}$ is the Courant factor. $c_m$ is the sound speed without radiation and $c_g$ is the contribution to the sound speed from each group

$$c_m = \sqrt{\frac{\gamma p}{\rho}}, \qquad c_g = \sqrt{\left(\frac{3-f_g}{2}\right)\left(\frac{\lambda_g E_g}{\rho}\right)}.$$

where $\gamma$ is the adiabatic index of the matter.

The parabolic component of the governing Eqs. (1-4) includes radiation diffusion, electron heat conduction, and source-sink terms. These terms, omitted from the hyperbolic subsystem discussion, are as follows:

$$\frac{\partial(\rho e)}{\partial t} = Q_L + \nabla \cdot (-\boldsymbol{F}_e) + \sum_{g=1}^{G} c\left(\kappa_{Pa,g} E_g - \kappa_{Pe,g} B_g\right), \tag{21}$$

$$\frac{\partial E_g}{\partial t} = \nabla \cdot \left(-\boldsymbol{F}_g\right) + \left(c\kappa_{Pe,g} B_g - c\kappa_{Pa,g} E_g\right). \tag{22}$$

Here, $e$ represents the specific internal energy. The term $c\left(\kappa_{Pe,g} B_g - \kappa_{Pa,g} E_g\right)$ symbolizes the energy exchange between matter and the $g$-th radiation group through absorption and emission of radiation.

The resolution of Eqs. (21) and (22) involves two stages. Initially, we implicitly update the radiation energy diffusion and the energy exchange between matter and each radiation group $g$. This is followed by a direct implicit update for electron heat conduction. These updates are critical for numerical stability due to the significantly shorter timescales of radiation energy diffusion, energy exchange, and heat conduction compared to the dynamical timescale.

The first implicit update, using the first-order backward Euler method, is formulated as follows:

$$\frac{\rho e^{**} - \rho e^{*}}{\Delta t} = \sum_{g=1}^{G} c\left(\kappa_{Pa,g}^{*} E_g^{**} - \kappa_{Pe,g}^{*} B_g^{**}\right), \tag{23}$$

$$\frac{E_g^{**} - E_g^{*}}{\Delta t} = \nabla \cdot D_g^{*} \nabla E_g^{**} + c\left(\kappa_{Pe,g}^{*} B_g^{**} - \kappa_{Pa,g}^{*} E_g^{**}\right). \tag{24}$$

In this context, the $*$ superscript indicates the first intermediate state following the explicit update, and the $**$ superscript denotes the second intermediate state after the first implicit update. The group Planck mean coefficients $\kappa_{Pe,g}^{*}$ and $\kappa_{Pa,g}^{*}$, as well as the diffusion coefficients $D_g^{*}$ are fixed at the time level $*$ (frozen coefficients). We exclude the $*$ superscript for $\rho^{*}$ as the implicit update does not alter the mass density.

Equations (23) and (24) are iteratively solved using the Newton-Raphson method. Initially, we define:

$$F_e = \rho e^{**} - \rho e^{*} - \Delta t \sum_{g=1}^{G} c\left(\kappa_{Pa,g}^{*} E_g^{**} - \kappa_{Pe,g}^{*} B_g^{**}\right), \tag{25}$$

$$F_g = E_g^{**} - E_g^{*} - \Delta t\left[\nabla \cdot D_g^{*} \nabla E_g^{**} + c\left(\kappa_{Pe,g}^{*} B_g^{**} - \kappa_{Pa,g}^{*} E_g^{**}\right)\right]. \tag{26}$$

The desired solution is for $F_e$ and $F_g$ to both be zero. We approach the solution by the Newton iteration, and in each iteration step we solve the following linear system:

$$\begin{bmatrix} (\partial F_e/\partial T)^{(k)} & (\partial F_e/\partial E_g)^{(k)} \\ (\partial F_g/\partial T)^{(k)} & (\partial F_g/\partial E_g)^{(k)} \end{bmatrix} \begin{bmatrix} \delta T^{(k+1)} \\ \delta E_g^{(k+1)} \end{bmatrix} = \begin{bmatrix} -F_e^{(k)} \\ -F_g^{(k)} \end{bmatrix}. \tag{27}$$

Here, $\delta T^{(k+1)} = T^{**,(k+1)} - T^{**,(k)}$ and $\delta E_g^{(k+1)} = E_g^{**,(k+1)} - E_g^{**,(k)}$, where the $(k)$ superscript denotes the stage of the Newton iteration. To reduce clutter, we drop the $**$ superscript without losing clarity.

Subsequently, $\delta T^{(k+1)}$ is eliminated from the system in Eq. (27), resulting in the following equation for $E_g^{(k+1)}$:

$$\left(\frac{1}{\Delta t} + c\kappa^*_{Pa,g}\right) E_g^{(k+1)} - \nabla \cdot D_g^* \nabla E_g^{(k+1)} = c\kappa^*_{Pe,g} B_g^{(k)} + \frac{E_g^*}{\Delta t} - \frac{\eta_g}{\Delta t}(\rho e^{(k)} - \rho e^*) +$$
$$\eta_g \sum_{g'} c \left(\kappa^*_{Pa,g'} E_{g'}^{(k+1)} - \kappa^*_{Pe,g'} B_{g'}^{(k)}\right). \tag{28}$$

where $\eta_g = -(\partial F_g/\partial T)(\partial F_e/\partial T)^{-1}$, the subscript $g'$ denotes energy groups distinct from the already used subscript $g$.

Equation (28) represents a set of coupled equations, interconnected through the summation terms on the right-hand side. To solve this system, we employ a strategy [43] involving decoupling and applying an iterative method, which is different from the Newton iteration used for Eqs. (23) and (24). This iterative process functions as an inner iteration, embedded within the outer Newton iteration step. Decoupling of Eq. (28) is achieved by substituting the radiative energy density in the right-hand side coupling term with the state from the preceding inner iteration. Each step of this inner iteration is governed by the following equation:

$$\left(\frac{1}{\Delta t} + c\kappa^*_{Pa,g}\right) E_g^{(k+1),l+1} - \nabla \cdot D_g^* \nabla E_g^{(k+1),l+1} = c\kappa^*_{Pe,g} B_g^{(k)} + \frac{E_g^*}{\Delta t} - \frac{\eta_g}{\Delta t}(\rho e^{(k)} - \rho e^*) +$$
$$\eta_g \sum_{g'} c \left(\kappa^*_{Pa,g'} E_{g'}^{(k+1),l} - \kappa^*_{Pe,g'} B_{g'}^{(k)}\right). \tag{29}$$

where $l$ denotes the inner iteration index. The inner iteration terminates when the maximum of the ratio $\sum_g \left|E_g^{(k+1),l+1} - E_g^{(k+1),l}\right|/\sum_g E_g^{(k+1),l+1}$ falls below a predetermined tolerance, such as $10^{-6}$.

It is critical to acknowledge that this inner iteration method may converge slowly in scenarios characterized by strong coupling (e.g., large opacities and small heat capacities) and/or substantial time steps, as discussed in [43]. To improve the convergence rate, we incorporate a local acceleration scheme as proposed by Zhang *et al.* [43]. Initially, we define the error in the radiation energy density, $\epsilon_g^{l+1}$, after $l$ inner iterations as:

$$\epsilon_g^{l+1} = E_g^e - E_g^{l+1}. \tag{30}$$

In this context, $E_g^e$ denotes the exact solution of Eq. (28), and $E_g^{l+1}$ signifies the result after the $l$th inner iteration as per Eq. (29). By subtracting Eq. (29) from Eq. (28), we derive the following equation for the error in the $g$th group:

$$\left(\frac{1}{c\Delta t} + \kappa_g\right) \epsilon_g^{l+1} - \eta_g \sum_{g'} \kappa_{g'} \epsilon_{g'}^{l+1} = \eta_g \sum_{g'} \kappa_{g'} \left(E_{g'}^{l+1} - E_{g'}^{l}\right). \tag{31}$$

Equation (31) can be analytically solved, yielding:

$$\epsilon_g^{l+1} = \left[\eta_g \sum_{g'} \kappa_{g'} \left(E_{g'}^{l+1} - E_{g'}^{l}\right)\right] \bigg/ \left[\left(\frac{1}{c\Delta t} + \kappa_g\right)\left(1 - \sum_g \frac{\eta_g \kappa_g}{\left(\frac{1}{c\Delta t} + \kappa_g\right)}\right)\right]. \tag{32}$$

Ultimately, we update the solution for each step of the inner iteration as follows:

$$E_g^{l+1} = E_g^{l+1} + \epsilon_g^{l+1}. \tag{33}$$

This acceleration process is executed post the convergence check of the inner iteration, ensuring that acceleration is only applied to solutions that have not yet converged.

Upon obtaining the numerical solution to Eq. (28) via the inner iteration, the matter internal energy density is updated as follows:

$$\rho e^{(k+1)} = \eta \rho e^{(k)} + (1-\eta)\rho e^* + (1-\eta)c\Delta t \sum_{g=1}^{G} \left( \kappa_{Pa,g}^* E_g^{(k+1),l+1} - \kappa_{Pe,g}^* B_g^{(k)} \right). \quad (34)$$

where $\eta = \sum_{g=1}^{G} \eta_g$. The new temperature $T^{(k+1)}$ is then derived from $\rho e^{(k+1)}$ and $\rho$ through a call to the EOS. Convergence of the outer iteration is determined by ensuring the relative temperature change between iterations falls below a user-specified tolerance, typically $10^{-6}$.

After the initial implicit update, the variables are obtained as $\boldsymbol{U}^{**} = \left( \rho^{n+1}, (\rho \boldsymbol{u})^{n+1}, (\rho E)^{**}, E_g^{n+1} \right)^T$. The subsequent step involves updating the contribution of heat conduction to the internal energy density of the matter in Eq. (21). This is accomplished by solving the equation:

$$\rho c_v \frac{\partial T}{\partial t} = \nabla \cdot C_e \nabla T + Q_L. \quad (35)$$

where $\rho c_v T = \rho e$, $c_v$ is the specific heat capacity of the matter. This equation is implicitly advanced as:

$$\rho c_v^{**} \frac{T^{n+1} - T^{**}}{\Delta t} = \theta (\nabla \cdot C_e^{**} \nabla T^{n+1}) + (1-\theta)(\nabla \cdot C_e^{**} \nabla T^{**}) + Q_L^{**}. \quad (36)$$

Here, the superscript $n+1$ indicates the final state at time $t^{n+1}$, subsequent to the second implicit update. $\theta = 1$ and $\theta = 0.5$ represent the Backward Euler and Crank Nicholson schemes, respectively. The source term $Q_L^{**}$ due to laser energy deposition is calculated based on $\rho^{**}$ and $T^{**}$ provided in the first implicit step using the laser deposition algorithm in Section 3.4. Following this second implicit update, the variables at time $t^{n+1}$ are determined as $\boldsymbol{U}^{n+1} = \left( \rho^{n+1}, (\rho \boldsymbol{u})^{n+1}, (\rho E)^{n+1}, E_g^{n+1} \right)^T$.

It should also be noted that, as outlined in Section 2, the RHDLPP-LTP package requires solving a specific heat conduction equation for the target, essential for modeling the laser energy transfer from the surface to the interior of the target. The governing equation is:

$$\rho_s c_v \left( \frac{\partial T_s}{\partial t} + v_c \frac{\partial T_s}{\partial z} \right) = \nabla \cdot C_e \nabla T_s + Q(z). \quad (37)$$

where $T_s$ and $\rho_s$ represent the temperature and mass density, respectively, with the subscript $s$ indicating the solid or liquid phase of the target. The term $v_c$ refers to the surface recession velocity. The variable $z$ specifies the direction perpendicular to the target surface, with $z_0$ marking the target surface. $Q(z) = S_{las}(t, z_0) \exp(-\alpha z)$, the laser energy deposited on the target surface is denoted by $S_{las}$, and $\alpha$ is the absorption coefficient. The equation is solved using the following implicit format similar to Eq. (36)

$$\rho_s^n c_v^n \frac{T_s^{n+1} - T_s^n}{\Delta t} = \theta \left( \nabla \cdot C_e^n \nabla T_s^{n+1} - v_c^n \frac{\partial T_s^{n+1}}{\partial z} \right) + (1-\theta) \left( \nabla \cdot C_e^n \nabla T_s^n - v_c^n \frac{\partial T_s^n}{\partial z} \right) + Q^n. \quad (38)$$

We convert the linear system Eqs. (29), (36), and (38) all into the following canonical form [43]

$$Af + \nabla \cdot B \nabla f + C \nabla f = RHS. \quad (39)$$

where $f$ is the variable to be solved, corresponding to $E_g^{(k+1),l+1}$ in Eq. (29), $T^{n+1}$ in Eq. (36) and $T_s^{n+1}$ in Eq. (38), respectively. $A$, $B$, $C$ are parameters with spatial variation. In 2D Cartesian coordinates, the term $\nabla \cdot B \nabla f$ is discretized using the following five-point stencil:

$$\nabla \cdot B \nabla f = \frac{\overline{B}_{i+1/2,j}(f_{i+1,j} - f_{i,j}) - \overline{B}_{i-1/2,j}(f_{i,j} - f_{i-1,j})}{\Delta x^2}$$
$$+ \frac{\overline{B}_{i,j+1/2}(f_{i,j+1} - f_{i,j}) - \overline{B}_{i,j-1/2}(f_{i,j} - f_{i,j-1})}{\Delta y^2}.$$

where $\overline{B}$ denotes the harmonic mean of the coefficients in neighboring cells. For example, the coefficient between cells $(i-1,j)$ and $(i,j)$ is defined as [51]

$$\overline{B}_{i-1/2,j} = \frac{2 B_{i,j} B_{i-1,j}}{B_{i,j} + B_{i-1,j}}.$$

Equation (39) is applicable to Cartesian, cylindrical, and spherical coordinates, provided that the appropriate metric factors are incorporated into the coefficients and the $RHS$ term. For solving this linear system, the RHDLPP employs the Lawrence Livermore National Laboratory's HYPRE package [52].

### 3.3. Equations of State

The RHD equations (1-4) require supplementation by EOS. The current code integrates two variants of EOS solvers. The first is based on a real gas approximation [35], appropriate for plasma characterized by lower temperatures and electron densities as produced by moderate-intensity ns-lasers in RHDLPP-LTP package. The second solver is a Quotidian Equation of State (QEOS) [36], which is suitable for LPPs exhibiting higher temperatures and densities in RHDLPP-HTP package. Subsequently, we shall discuss each of these EOS solvers in detail.

(1) **EOS based on a real gas approximation**: As introduced in Section 2, there will be a jump between the target parameters and the plasma parameters due to the existence of the KL when the moderate-intensity ns-laser acts. This phenomenon enables separate simulation of target material and plasma within the RHDLPP-LTP package. Therefore, the EOS only needs to be fully applicable to the plasma state. Because the contribution of excitation energy to internal energy is very important in the early stage of hot plasma evolution, and the ionization energies of atoms even ions have an important influence on the plasma temperature, density and emission spectrum for a long time (in the order of microsecond) in the evolution of the plasma [53]. So the EOS based on a real gas approximation, which divides the internal energy into the thermal energy of atoms, ions and electrons, ionization energy and the excitation energy of atoms as well as ions, is established [35]:

$$\rho e = \frac{3}{2}(1+\langle z \rangle)nkT + n \sum_s f_s \sum_{z=1} f_{z,s} \sum_{l=0}^{z-1}(E_{l,s} - \Delta E) + n \sum_s f_s \sum_{z=0} f_{z,s} \sum_{j>j_0} f_{j,z,s} E_{j,z,s}, \quad (40)$$

$$p = kT(n_e + n), \quad (41)$$

$$f_s = \frac{n_s}{n}, \quad f_{z,s} = \frac{n_{z,s}}{n_s}, \quad f_{j,z,s} = \frac{n_{j,z,s}}{n_{z,s}}, \quad \langle z \rangle = \sum_s f_s \sum_{z=1} z f_{z,s}. \quad (42)$$

where $n$ is the total atomic number density, $n = \sum_s \sum_z n_{z,s}$. $n_s$ refer to the total number density of atomic species $s$, $n_s = \sum_z n_{z,s}$. $n_{z,s}$ and $E_{z,s}$ refer to the number density and ionization energy of $zth$ ionization state of gas species $s$, respectively. $z = 0$ corresponds to atoms. $\Delta E$ is a correction to the ionization energy and calculated by Debye shielding. $j_0$ represents the ground state of each ion. $E_{j,z,s}$ is the excitation energies of the energy level $j$ with respect to the ground state energy. $f_s$, $f_{z,s}$, $f_{j,z,s}$, and $\langle z \rangle$ represent the relative species fractions, ionization fractions, excitation fractions, and average charge state.

(2) **QEOS**: When a target is irradiated by an intense laser pulse, the KL between the condensed and plasma phases vanishes, transforming into a macroscopic transition layer. Consequently, the RHD equations must be resolved across the entire physical domain, including the transition layer across which the density changes very rapidly, the high-density condensed phase below the transition layer, and the low-density and high-temperature plasma above the transition layer. Therefore, a EOS capable of encompassing a very wide range of temperature and density is required. In RHDLPP-LTP package, the QEOS [36], which satisfies well the above requirement, is employed.

In our QEOS, thermodynamic functions like pressure, energy, entropy, and Helmholtz free energy are calculated as a sum of the zero temperature isotherm and thermal contributions from ions and electrons. The electronic contribution to the total EOS is calculated from the Thomas-Fermi (TF) model in its simplest version without exchange and gradient corrections. The TF model at finite temperatures follows from the Poisson equation for the self-consistent atomic potential $V(r)$, and a boundary problem can be defined as [54]

$$\frac{1}{r}\frac{d^2}{dr^2}(rV) = \frac{2}{\pi}(2T)^{3/2} I_{1/2}\left(\frac{V(r)+\mu}{T}\right), \tag{43}$$

$$rV(r)|_{r=0} = Z, \quad V(r_0) = 0, \quad \left.\frac{dV(r)}{dr}\right|_{r=r_0} = 0. \tag{44}$$

Here we use the atomic system of units. $r$ is the radial coordinate, $\mu$ is the chemical potential, $I_k(x)$ represents the Fermi-Dirac integral. $Z$ is the nuclear charge, $r_0$ denotes the radius of the ion spherical cell.

The specific internal energy $E_e$ and pressure $P_e$ of electron are given by [54]

$$P_e = \frac{(2T)^{5/2}}{6\pi^2} I_{3/2}(\phi(1)), \quad E_e = \frac{\sqrt{2}vT^{5/2}}{\pi^2}\left[2I_{3/2}(\phi(1)) - 3\int_0^1 I_{3/2}\left(\frac{\phi(x)}{x}\right)x^2 dx\right] - E_0. \tag{45}$$

where $x$ is defined as $0 < x = r/r_0 \leq 1$, and the variable $\phi(x)$ is defined as $\phi(x)/x = (V(r)+\mu)/T$. $v$ represents the volume related to one ion. $E_0 = -0.76874512422$ is the internal energy of an isolated atom.

After some mathematical manipulation [55], Eqs. (43)-(45) can be reduced to the following Cauchy problem [55]

$$\phi'(x) = \xi(x), \tag{46}$$

$$\xi'(x) = ax I_{1/2}\left(\frac{\phi(x)}{x}\right), \tag{47}$$

$$\phi(1) = \xi(1), \quad \phi(0) = \frac{Z}{Tr_0}, \tag{48}$$

$$P_e = \frac{(2T)^{5/2}}{6\pi^2} I_{3/2}(\phi(1)), \tag{49}$$

$$E_e'(x) = \frac{3\sqrt{2}vT^{5/2}}{\pi^2} I_{3/2}\left(\frac{\phi(x)}{x}\right) x^2, \tag{50}$$

$$E_e(1) = \frac{2\sqrt{2}vT^{5/2}}{\pi^2} I_{3/2}(\phi(1)) - E_0. \tag{51}$$

where $a = 4\sqrt{2T}r_0^2/\pi$. The procedures to solve the Cauchy problem are: first, use the shooting method to solve Eqs. (46)-(48) to get $\phi(1)$, and then substitute $\phi(1)$ into Eq. (49) to get $P_e$. Finally, the adaptive stepsize Runge-Kutta integration [56] is used to solve Eqs. (50) and (51) to obtain the value of energy $E_e \equiv E_e(0)$.

The energy and pressure in a solid at zero temperature (the zero temperature isotherm) are determined by a scaled binding energy (SBE) model [57, 58]. The SBE model defines the density-dependent cold curve quantities for the energy $E_c$, pressure $P_c$, and bulk modulus $B$ in terms of the scaled binding energy function $\varepsilon(a)$ and its derivatives. The energy $E_c$ is represented as

$$E_c = E_{coh}[1 + \varepsilon(a)], \quad a = \frac{r - r_0}{l}, \quad l = \left[\frac{E_{coh}Am_p}{12\pi B_0 r_0}\right]^{1/2}. \tag{52}$$

where $E_{coh}$ is the cohesive energy. $r$ is the Wigner-Seitz (WS) cell radius based on the current number density of the material, and $r_0$ is the WS cell radius corresponds to the normal density $\rho_0$. $l$ represents the scale length, $A$ denotes the atomic weight, $m_p$ is the mass of proton, and $B_0$ is the bulk modulus at normal density. The functional form of $\varepsilon(a)$ is of the form

$$\varepsilon(a) = \frac{e^{-a}}{R^2}[\alpha + \beta a + \gamma a^2 + \lambda a^3]. \tag{53}$$

where $R = 1 + ka$ and $k = l/r_0$. The constraints for the function $\varepsilon(a)$ are $\varepsilon(0) = -1$, $\varepsilon'(0) = 0$, $\varepsilon''(0) = 1$, and $\varepsilon'''(0) = -3k(B_0' - 1)$. $B_0'$ is the bulk modulus pressure derivative, its value is given by experiments or theoretical calculations. In order for $\varepsilon(a)$ and its derivatives to be numerically stable, $B_0'$ must satisfy the inequality $B_0' > 1 + 2/(3k)$ [58]. By utilizing these constraints, the coefficients in function $\varepsilon(a)$ can be obtained as $\alpha = -1$, $\beta = -1 - 2k$, $\gamma = -2k - k^2$, and $\lambda = 1/3 - k(B_0' - 1)/2 - k^2$.

The pressure $P_c$ in a solid at zero temperature and the cold curve bulk modulus $B$ are given by

$$P_c = -3kB_0 \left[\frac{\varepsilon'(a)}{(1 + ka)^2}\right], \quad B = B_0 \left[\frac{\varepsilon''(a)}{1 + ka} - \frac{2k\varepsilon'(a)}{(1 + ka)^2}\right]. \tag{54}$$

The thermodynamic contribution of the ions to the total EOS is calculated by using the Cowan model [36], which employs analytical formulas to smoothly interpolate between the Debye solid, the normal solid, and the liquid states. The Cowan model consists of two independent parts. The phenomenological part of the model makes estimations for the Debye temperatures $\Theta_D$ and the melting temperatures $T_m$ depending on the density. In the structural part, the pressure $P_i$, energy $E_i$, entropy $S_i$, and Helmholtz free energy $F_i$ are calculated.

In the phenomenological part, we still use the SBE model to calculate the Debye and melting temperatures. The detailed derivation process can be seen in Ref. [58], and in this paper, we only provide the final expression for the density-dependent melting temperature $T_m(a)$ and the density-dependent Debye temperature $\Theta_D(a)$ as follows

$$T_m(a) = \begin{cases} T_{m0}(1+ka)[(1+ka)\varepsilon''(a) + 2k(t-1)\varepsilon'(a)] & \text{if } a \leq 0 \\ T_{m0}(1+ka)e^{-ak(6\gamma_0-1)} & \text{otherwise} \end{cases}, \quad (55)$$

$$\Theta_D(a) = \begin{cases} \Theta_{D0}\sqrt{\varepsilon''(a) + \dfrac{2k(t-1)\varepsilon'(a)}{1+ka}} & \text{if } a \leq 0 \\ \Theta_{D0}\sqrt{\dfrac{e^{-ak(6\gamma_0-1)}}{1+ka}} & \text{otherwise} \end{cases}. \quad (56)$$

where $T_{m0}$ and $\Theta_{D0}$ are the melting and Debye temperatures of the solid at zero pressure, respecyively. $\gamma_0$ denotes the Grüneisen coefficient of the solid phase of the material at zero pressure. $t = 2.5 - 3(1 + \gamma_0 - B_0'/2)(1+ka)$ represents the Grüneisen scaling parameter.

In the structural part, we define the scaling variables, $u = \Theta_D(a)/T$, and $w = T_m(a)/T$, with reference to a known melting temperature and Debye temperature. The material is considered to be in the solid phase when $w$ is greater than one and in the fluid phase (liquid, gas or plasma) when $w$ is less than one. In the Cowan model there are two classes of solid material, low-temperature solids ($u > 3$) and high-temperature solids ($u < 3$). The Cowan model defines a scaling function $f(u,w)$ as follows

$$f(u,w) = \begin{cases} -\dfrac{11}{2} + \dfrac{9}{2}w^{1/3} + \dfrac{3}{2}\log\left(\dfrac{u^2}{w}\right), & w < 1 \\ -1 + 3\log u + \left(\dfrac{3u^2}{40} - \dfrac{u^4}{2240}\right), & w > 1, u < 3 \\ \dfrac{9u}{8} + 3\log(1-e^{-u}) - \dfrac{\pi^4}{5u^3} + e^{-u}\left(3 + \dfrac{9}{u} + \dfrac{18}{u^2} + \dfrac{18}{u^3}\right). & w > 1, u > 3 \end{cases} \quad (57)$$

The scaling function $f(u,w)$ and its derivative $f'(u,w)$ are used to calculate the ion Helmholtz free energy $F_i$, ion internal energy $E_i$, ion entropy $S_i$, and ion pressure $P_i$ as follows

$$F_i = Tf(u,w), \quad (58)$$

$$E_i = Tuf'(u,w), \quad (59)$$

$$S_i = uf'(u,w) - f(u,w), \quad (60)$$

$$P_i = n_i \begin{cases} \gamma_s E_i, & w \geq 1, \\ (1 + (3\gamma_s - 1)w^{1/3})T, & w < 1. \end{cases} \quad (61)$$

where $\gamma_s$ represents the Grüneisen parameter and its specific expression can be found in Ref. [58].

Moreover, our QEOS model can describe the liquid-vapor phase coexistence through a fully equilibrium EOS which is obtained by a Maxwell construction eliminating the van-der-Waals loops. The specific implementation is completely consistent with that in the "Frankfurt equation-of-state (FEOS)" package [59], and will not be explained in detail here.

3.4. Average ionization degree

Although the average ionization degrees are not involved in solving the multigroup radiation hydrodynamics equations in Section 3.1, they are necessary physical parameters in the laser energy deposition calculation in Section 3.4 below. In RHDLPP, the energy level population, charge state distribution and average ionization degree are calculated by a steady-state CR model including various atomic processes based on screened-hydrogenic approximation [38]. Under the steady-state approximation, the CR model needs to establish and solve the following rate equations

$$\sum_{j=1}^{j=i-1} R_{j\to i} n_{j,z,s} + \sum_{j=i+1}^{m} R_{j\to i} n_{j,z,s} - R_i n_{i,z,s} = 0. \tag{62}$$

where $R_{j\to i}$ and $R_i$ represent the rates from the $jth$ energy level to the $ith$ energy level and the loss rate of the $ith$ energy level, respectively.

**Table 1**. List of atomic processes included in RHDLPP and the semiempirical formulas used to calculate the corresponding rates.

| Atomic process | Rate coefficient | Expression |
|---|---|---|
| Collisional excitation | $\mathcal{E}_{Z,i\to Z,j}$ | Van Regemorter |
| Collisional deexcitation | $\mathcal{D}_{Z,j\to Z,i}$ | Detailed balance |
| Collisional ionization | $\mathcal{I}_{Z-1,i\to Z,j}$ | Lotz |
| Three-body recombination | $\mathcal{R}^3_{Z+1,i\to Z,j}$ | Detailed balance |
| Photo-absorption | $\mathcal{P}a_{Z,i\to Z,j}$ | DEDALE approach [60] |
| Photo-emission | $\mathcal{P}e_{Z,j\to Z,i}$ | Detailed balance |
| Photo-ionization | $\mathcal{P}i_{Z-1,i\to Z,j}$ | Kramers |
| Photo-recombination | $\mathcal{P}r_{Z+1,i\to Z,j}$ | Detailed balance |
| Autoionization | $\mathcal{A}u_{Z-1,i\to Z,j}$ | FLYCHK approach [38] |
| Electron capture | $\mathcal{C}_{Z+1,i\to Z,j}$ | Detailed balance |

We considered the following 12 atomic processes in the CR model: collisional ionization and three-body recombination, collisional excitation and deexcitation, photo-absorption, spontaneous and stimulated emission, photo-ionization, spontaneous and stimulated recombination, autoionization, and electron capture, which are summarized in Table 1.

3.5. Laser propagation and energy deposition

As the only external source term, the propagation and energy deposition of incident laser in the target and plasma are important factors determining the characteristics of LPP. In the code, a hybrid model [34] that combines the geometrical-optics ray-tracing method in the sub-critical plasma regions with the 1D solution of the Helmholtz wave equation in the super-critical areas is used to address the issue of laser propagation and energy deposition.

The solution process is mainly divided into the following steps: Firstly, the incident laser beam is split into a large number of infinitely thin parallel rays along the cross-section of the beam, which have different powers due to the radial intensity distribution of the incident laser beam. Subsequently, each ray is processed using the geometric optics approach [61] when propagated in an underdense region where the electron density $n_e$ is less than the critical density $n_c$ of the plasma. There the trajectory of a ray is calculated by the ray equation of motion [61]

$$\frac{d^2\vec{r}}{dt^2} = \nabla\left(-\frac{c^2}{2}\frac{n_e}{n_c}\right). \tag{63}$$

where $c$ is the speed of light in vacuum and $n_c = (m_e/4\pi)(\omega/e)^2$. $\omega$ is the laser frequency, the quantities $m_e$ and $e$ are the electron mass and charge, respectively. The attenuation of power $P_r(t)$ during ray propagation is dominated by the inverse Bremsstrahlung process and can be calculated by a first-order ordinary differential equation [61]

$$\frac{dP_r}{dt} = -\nu_{ib}(t)P_r. \tag{64}$$

where $\nu_{ib} = (n_e/n_c)\nu_{ei}$ is the inverse Bremsstrahlung frequency factor and $\nu_{ei}$ is the electron-ion collision frequency.

Thirdly, when the rays propagate to the vicinity of the critical density interface, they will be split into reflected rays (*rr*) that return to the underdense region and transmission rays (*tr*) that enter the overdense region. The trajectory tracking and energy deposition of *rr* in the underdense region are still achieved by those as mentioned above geometric optical methods, while the energy deposition of *tr* in the overdense area is implemented by solving the 1D Helmholtz equation of the *s*- and *p*-polarized components of the plane wave as follows [19, 34]

$$s: \quad \frac{d^2 E}{dz^2} + \frac{\omega^2}{c^2}(\varepsilon - \alpha^2)E = 0, \quad \frac{dE}{dz} = i\frac{\omega}{c}H, \tag{65}$$

$$p: \quad \frac{d^2 H}{dz^2} + i\frac{\omega E}{c}\frac{d\varepsilon}{dz} + \frac{\omega^2}{c^2}(\varepsilon - \alpha^2)H = 0, \quad \frac{dH}{dz} = -i\frac{\omega\varepsilon}{c}E. \tag{66}$$

where $E$ and $H$ are the complex amplitudes of the electric field and magnetic field, respectively. $\omega$ represents the angular frequency of the laser, and $\alpha = k_\perp c/\omega$ with the transversal component $k_\perp$ of the wavenumber. In vacuum regions, $\alpha$ reduces to the $sin\theta$ where $\theta$ is the angle of incidence. $\varepsilon$ represents the complex dielectric permittivity of the medium, defined as $\varepsilon = 1 - \omega_p^2/(\omega^2 + i\omega\nu_{ei})$. $\omega_p = \sqrt{4\pi e^2 n_e/m_e}$ is the plasma frequency. The solution of this 1D wave optics problem yields the reflected, transmitted, and absorbed fractions of the incident energy flux. The reflection fraction corresponds to the power of the *rr*, while the transmission and absorption fractions correspond to the deposition of the *tr* in each cell of the overdense region and the transmittance at the cell surface.

The ray equation (63) and the power deposited equation (64) under geometric optics approximation are numerically solved using the laser propagation-deposition scheme developed by Kaiser [61]. Consider the piecewise constant approximation of the function that describes the complex dielectric permittivity of the medium [19, 34], the 1D Helmholtz wave equations (65) and (66) are numerically solved using the finite-analytic method [19, 34]. A special rescaling procedure of the wave optics deposition profile proposed in Ref. [34] was used to ensure a smooth match between the geometrical optics and wave optics deposition powers at the transition point. For more detailed information about the solution procedure, please refer to Refs. [19], [34] and [62].

Our code incorporates two different models of the electron-ion collision frequency. For the hot plasma, the collision frequency is given by Spitzer's formula [19]:

$$\nu_{ei} = \frac{4\sqrt{2\pi}e^4\langle z\rangle n_e}{3\sqrt{m_e}(kT_e)^{3/2}}\ln(\Lambda). \tag{67}$$

Here, $\langle z \rangle$ represent the average ionization degree, and $\ln(\Lambda)$ is the Coulomb logarithm, where $\Lambda = 1.0 + \max(1.0, b_{max}/b_{min})$ . The maximum collision parameter is $b_{max} = \sqrt{kT_e/m_e}/\max(\omega, \omega_p)$ and the minimum collision parameter $b_{min} = \max[\langle z\rangle e^2/(2kT_e), \hbar/(2\sqrt{kT_e m_e})]$.

Alternatively, for warm dense matter and cold solids, an ad hoc interpolation formula based on the Drude-Sommerfeld model is available [19, 63]:

$$\nu_{ei} \cong 2\sqrt{2\pi}\frac{e^4\langle z\rangle n_e}{\sqrt{m_e}(kT_e)^{3/2}}\ln\left[1.0 + K_{ds}\frac{1.32}{\sqrt{2\pi}}\frac{kT_e}{\left(\sqrt{m_e}e^2\langle z\rangle\widetilde{\omega}\right)^{2/3}}\right]F(T_e,\hbar\omega). \qquad (68)$$

where $\widetilde{\omega} = \max(\omega,\omega_p)$, $F(T_e,\hbar\omega) = \sqrt{\pi/2}\,\langle u_{th}/u\rangle$ represents the Fermi factor, and $\langle u_{th}/u\rangle$ denote averaging over the Fermi distribution function including Pauli blocking [19]. $K_{ds}$ is a free parameter and the default value is set to $K_{ds} = 1$. For $T_e \to \infty$, Eq. (64) becomes parametrically identical to the collision frequency Eq. (67) of a hot plasma; on the other hand, for $T_e \to 0$, the corresponding equations for warm dense matter and cold solids can then be derived.

3.6. Spectral post-processor of radiation hydrodynamics simulation

Spectral analysis and simulation are essential tools for diagnosing the states of LPPs and for validating the results of RHD codes. If the plasma has no significant spatial inhomogeneities and the radiation is transported in it with minimal absorption and scattering (indicating an optically thin plasma), results such as emissivities computed by 0D codes-like ATOMIC [64], THERMOS [54], and DLAYZ [65]-that model the population kinetics and radiative properties of plasmas can be compared with experimentally obtained spectra. However, the assumptions of homogeneity and optical thinning often do not hold for LPPs. To address these complexities, we have developed a post-processing code for spectral simulation within the RHDLPP code, called RHDLPP-SpeIma3D. This code eliminates the need for assumptions regarding plasma uniformity and the dominance of either emissivity or opacity in spectral formation. Consequently, it is now capable of simulating transient evolution images, temporally and spatially resolved spectra, and integral spectra across varying wavelength ranges that are directly comparable to experimental results.

Subsequently, we will employ the simulation of transient images in the 200-850 nm wavelength range as an exemplar to elucidate the simulation methodology [66] of the post-processing code. The schematic of transient image simulation is shown in Fig. 2. This method simulates the experimental images in three stages [66]. First, the distributions of plasma state parameters, such as temperature, density and ion abundance, are calculated using the RHDLPP code. Then, the 2D state parameter distributions are transformed into 3D distributions along the symmetry axis based on the axisymmetric properties of the plasma. The 3D plasma is encased by a bounding cylindrical surface, with the unperturbed background gas filling the space between them.

The second stage involves constructing ray paths based on an imaging system composed of a detector, an imaging lens, and a 3D plasma illumination source, i.e., ray tracing is executed. As shown in Fig. 2(a), the sensor surface is initially divided into multiple pixels. Following this, a symmetry plane parallel to the sensor surface is designated as the object plane within the plasma. The object plane is divided into an array of surface elements corresponding to the pixel array of the sensor, based on the object-image relationship of the imaging lens and the pixel size. Each surface element is denoted by the symbol $\varepsilon$. Subsequently, the lens surface ($\mathcal{L}$) is divided into equal-area cells, each denoted by the symbol $d\sigma$, while $\varepsilon$ on the object surface is divided into smaller cells denoted by $ds$. If the areas of $ds$ and $d\sigma$ are sufficiently small, ray paths can be constructed by connecting their geometric centers with a straight line. As illustrated in Fig. 2(a), each path terminates at the center of $d\sigma$ on the imaging lens, while the starting point moves from the center of $ds$ along the ray to the bounding cylindrical surface, ensuring that the ray traverses the plasma.

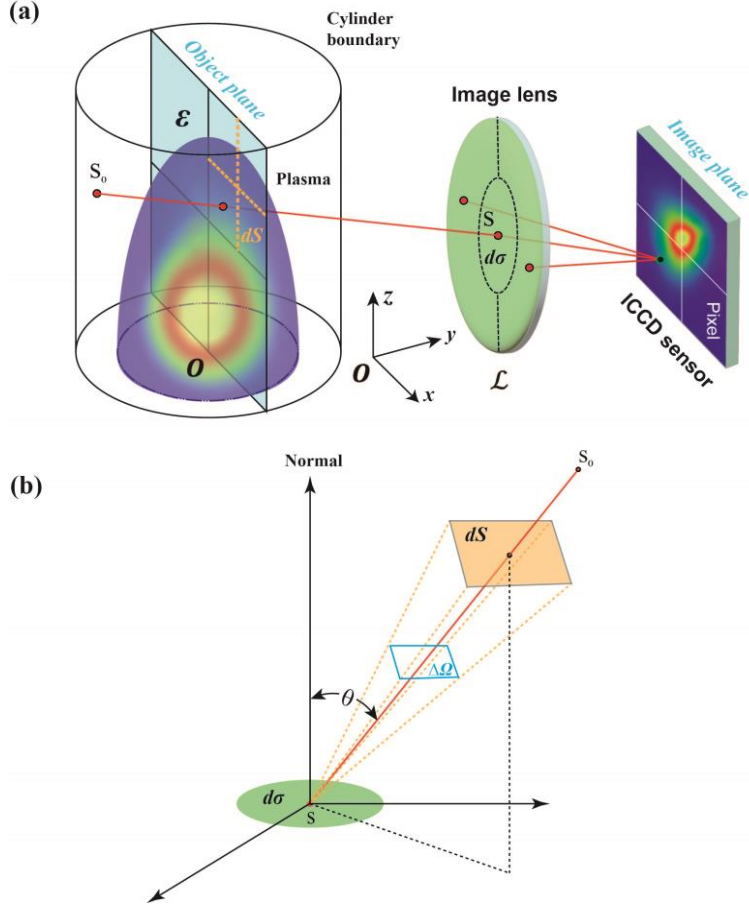

**Fig. 2.** Schematic diagram of the simulation reconstruction of the plasma transient image using RHDLPP-SpeIma3D code.

In the third step, the specific intensity $I_\nu$ of the radiation with frequency $\nu$ at the endpoint of the ray at any given moment is calculated. This is achieved by resolving the 1D integral form of the steady-state radiative transfer (RT) equation along each path [33]:

$$I_\nu = I_{\nu 0}\exp\left(-\int_{s_0}^{s}\kappa_\nu(s')\,ds'\right) + \int_{s_0}^{s} j_\nu(s')\left[\exp\left(-\int_{s'}^{s}\kappa_\nu(s'')\,ds''\right)\right]ds'. \qquad (69)$$

Here, $s_0$ represents the ray's starting point on the bounding cylindrical surface as depicted in Fig. 2(a), and $I_{\nu 0}$ signifies the radiation intensity at this location. Symbol $s$ denotes the endpoint of each path on the centroid of $d\sigma$ on the imaging lens, while $j_\nu$ and $\kappa_\nu$ denote the local emissivity and absorptivity in space and time, respectively.

The radiant energy, $E(ds \to d\sigma)$, at frequency $\nu$, collected by the cell $d\sigma$ from $ds$ at any given moment, is computed using the equation $E(ds \to d\sigma) = I_\nu \cdot \Delta S \cdot \Delta\Omega \cdot \cos\theta$. Here, $\Delta S$ represents the area of $d\sigma$, $\Delta\Omega$ denotes the solid angle subtended by $ds$ at the center of $d\sigma$, and $\theta$ is the angle between a ray and the normal of $d\sigma$, as illustrated in Fig. 2(b). This equation holds validity if the areas of $ds$ and $d\sigma$ are sufficiently small to ensure that the ray between their centers fully represents the rays between all points in the two cells. In the simulation, the adequacy of the areas is assessed through the convergence of the results. Subsequently, the radiant energy, $E(\varepsilon \to d\sigma)$, collected by $d\sigma$ from surface element $\varepsilon$, is determined by computing the radiant energy collected by $d\sigma$ from all $ds$ cells on the surface element $\varepsilon$ and summing them. The total radiant energy $E(\varepsilon \to Pixel)$ for each pixel on the sensor corresponding to $\varepsilon$ is calculated by summing up $E(\varepsilon \to d\sigma)$ collected by each $d\sigma$ cell on the lens $\mathcal{L}$ from the same $\varepsilon$. Finally, the number of counts per pixel within the time interval $\Delta t$ for the frequency range from $v$ to $v + \Delta v$ can be calculated using the integral equation:

$$N_{Counts} = \iint_{t,v}^{t+\Delta t, v+\Delta v} \frac{E(\varepsilon \rightarrow Pixel)}{hv} F(v)T(v)dvdt. \tag{70}$$

where $hv$ refers to the photon energy, $F(v)$ is the sensor's response function, and $T(v)$ represents the transmittance of the lens and optical filters. $\Delta v$ and $\Delta t$ are determined by the measured wavelength range and the gate width of the detector.

The simulation of transient images is accomplished by repeating the outlined operation for each pixel unit. To implement the simulation of spectra, the following modifications need to be made to the imaging simulation steps: (1) Replace the sensor's surface depicted in Fig. 2 with the plane where the entrance slit is located. (2) Partition this plane into several cells of equal area, similar to the sensor's pixels. (3) The establishment of a new object plane in the plasma is based on the area of the slit's plane and the inherent object-image relationship of the lens. (4) Once the object plane is divided twice, the ray path linking points on the object plane and the imaging lens is determined. (5) To determine the total radiant energy entering the slit, the radiant energy collected by each small cell on the slit plane is aggregated. This methodology enables the simulation and examination of temporally and spatially resolved or integrated spectra across the visible to infrared wavelength range that are relevant in LIBS. Moreover, this approach can be extended to simulate EUV spectra and conversion efficiencies that are pertinent to EUVL.

The radiative parameters of plasma, including emissivity and absorption coefficients in Eq. (69), along with Rosseland and Planck multigroup opacities in Eqs. (3) and (4), are predominantly obtained from the TOPS Opacities database [67] and the THERMOS database [54]. At present, we are developing a versatile, independent code to model the population kinetics and radiative properties of plasmas in a state of non-local thermodynamic equilibrium, utilizing a detailed configuration accounting approach. The advancements and outcomes of this development will be detailed in future publications.

3.7. Knudsen layer boundary conditions in RHDLPP-LTP package

In RHDLPP-LTP package, the RHD equations governing plasma dynamics and heat conduction equation for target material are coupled through the KL relation [37, 40, 68]. KL with a thickness of only a few mean free paths between the target surface and vapor plasma can be regarded as a discontinuity in the model, and the state parameters (such as density, temperature, and pressure) of the target surface and the vapor plasma adjacent to the KL may have a jump. To accurately represent this discontinuity, the RHDLPP-LTP package employs an analytical expression [37, 68] that effectively links the target surface with the vapor plasma in the vicinity of the KL

$$\frac{T_K}{T_s} = \left[\sqrt{1 + \pi\left(\frac{\gamma-1}{\gamma+1}\frac{m_0}{2}\right)^2} - \sqrt{\pi}\frac{\gamma-1}{\gamma+1}\frac{m_0}{2}\right]^2, \tag{71}$$

$$\frac{\rho_K}{\rho_s} = \sqrt{\frac{T_s}{T_K}}\left[(m_0^2 + 0.5)e^{m_0^2}erfc(m_0) - \frac{m_0}{\sqrt{\pi}}\right] + 0.5\frac{T_s}{T_K}\left[1 - \sqrt{\pi}m_0 e^{m_0^2}erfc(m_0)\right], \tag{72}$$

$$\rho_s v_c = p_s\left(\frac{m_a}{2\pi k T_s}\right)^{1/2} - \beta\rho_K\sqrt{\frac{R_v T_K}{2\pi}}\left[e^{-m_0^2} - \sqrt{\pi}m_0 erfc(m_0)\right], \tag{73}$$

$$\beta = \left[(2m_0{}^2 + 1) - m_0\sqrt{\frac{\pi T_s}{T_K}}\right] e^{m_0{}^2} \frac{\rho_s}{\rho_K}\sqrt{\frac{T_s}{T_K}}. \tag{74}$$

where $\rho_s$ and $T_s$ are the target surface density and temperature, respectively. $T_K$ and $\rho_K$ are the temperature and density of the vapor plasma near the KL, respectively. $R_v$ represents the gas constant, $\gamma$ denotes the ratio of specific heats, and $m_a$ represents the atomic mass. The reduced Mach number $m_0$ is related to the Mach number $M$ of the vapor plasma adjacent to the KL, $m_0 = \sqrt{\gamma/2}\, M$. The complementary error function $\text{erfc}(m_0)$ is given by $\text{erfc}(m_0) = \int_{m_0}^{\infty} e^{-\xi^2} d\xi$. $p_s$ represents the saturation vapor pressure of the target surface and can be calculated by

$$p_s = p_0 \exp\left[\frac{\Delta H_{lv}}{R_v}\left(\frac{1}{T_b} - \frac{1}{T_s}\right)\right], \quad p_0 = 1\,\text{atm}. \tag{75}$$

In addition, since there are no disturbances propagating from the gas to the boundary, supersonic evaporation of the target surface is impossible [40]. Therefore, the jump condition applied to the evaporation process should be restricted to $0 \leq M \leq 1$.

The saturated vapor pressure $p_s$ will drop down with the decreases of the surface temperature $T_s$, and may eventually be lower than the vapor pressure $p_K$ near the KL. At this time, back condensation ($M < 0$) begins to occur. In contrast to evaporation, condensation can be supersonic or subsonic. In the supersonic case, all boundary variables are extrapolated from the RHD equations, while in the subsonic case, only the following pressure ratio condition is required

$$\frac{p_K}{p_s} = 0.95 e^{(2.42|M|)}. \tag{76}$$

## 4. Code verification

In this section, we present detailed tests of the RHDLPP code demonstrating its ability to handle a wide range of radiation hydrodynamics problems.

### 4.1. The Reinicke & Meyer-ter-Vehn (RMTV) problem

The RMTV problem is a single-temperature hydrodynamics test including heat conduction, which can study the influence of nonlinear heat conduction on the Sedov-Taylor point explosion in a background gas [69]. The heat conductivity in this problem is a nonlinear function of temperature and density in power-law form, $C_{v,e} = \rho^a T^b$. We consider the spherically symmetric case, taking the coefficient $a = -2$, $b = 6.5$, and the adiabatic index is $\gamma = 5/4$. Reinicke and Meyer-ter-Vehn examined the spherically symmetric case and found that the hydrodynamics equations can be reduced to a set of four ordinary differential equations (ODEs) [69]. By evaluating the initial conditions and numerical integration of ODEs, self-similar analytical solutions for the point explosion with heat conduction can be derived.

We perform simulations in 1D spherical and 2D cylindrical (rz-geometry) coordinates. The computational domains in 1D and 2D simulations are divided in 400 and 400×400 cells, respectively. The spherical self-similar solution with the shock front located at the spherical radius of 0.225 and the heat front at 0.45 is used as the initial condition for the simulation. Fig. 3 shows the normalized temperature, density, and radial velocity profiles for 1D simulation at $t = 0.5125$ ns. The numerical simulation result shown by the circle symbols is close to the self-similar analytical solution shown by the solid lines. Except for the heat front located at 0.9, the temperature distribution is smooth due to the heat conduction. The slight deviation between the density and radial velocity near $r = 0.3$ and the analytical solution is due to the diffusion of the shock front in the initial conditions during the first few time steps. The 2D distributions of temperature and density at $t = 0.5125$ ns are shown in Figs. 4(a) and 4(c). Obviously, the Cartesian grid with $rz$-geometry does not significantly distort the spherical symmetry of the solution. The distributions of the relative error in the temperature and density (compared to the analytical solution) are shown in Figs. 4(b) and 4(d). As expected, the maximum error occurs at the discontinuity of the shock and heat front.

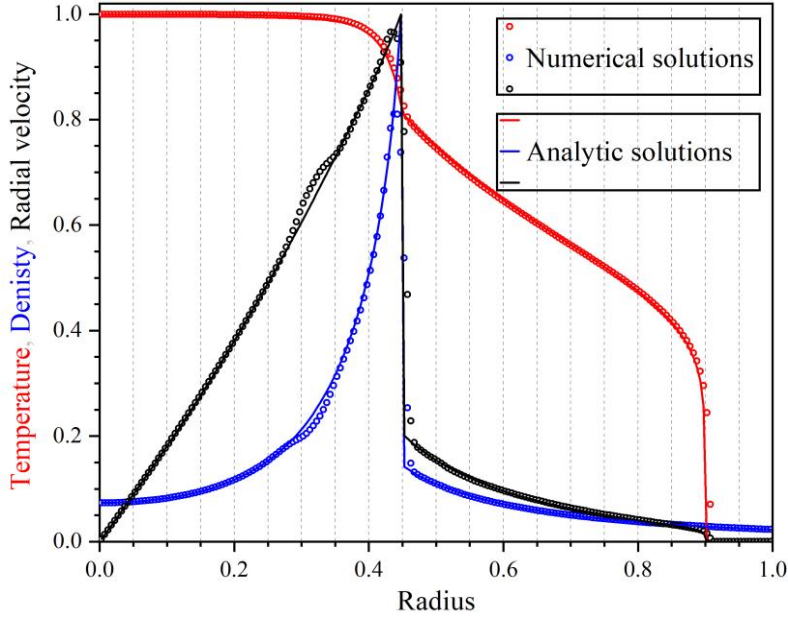

**Fig. 3.** Normalized temperature (red), density (blue) and radial velocity (black) profiles at 0.5125 ns for the RMTV problem. Numerical and analytic solutions are shown in circle symbols and solid lines, respectively.

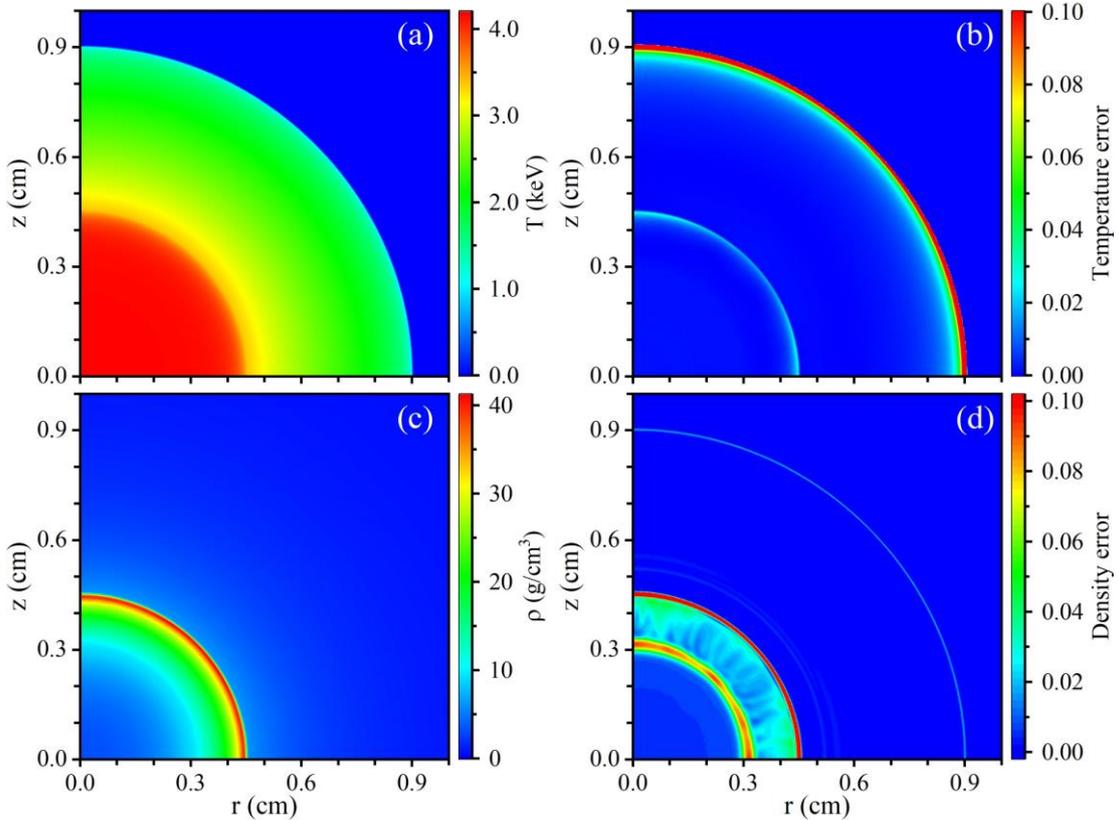

**Fig. 4.** (a) Temperature, (b) relative error in the temperature, (c) density, and (d) relative error in the density, for the RMTV problem on 400×400 grids in $rz$-geometry at 0.5125 ns.

## 4.2. Radiative blast wave

A blast wave is a type of propagating disturbance generated by the release of energy in a small volume of gas. In the case of an explosion, the initial release of energy creates a shock wave that moves outward, compressing and heating the gas it encounters. When the blast wave is so intense that it significantly heats the gas, it can cause the gas to emit radiation, creating what is known as a radiative blast wave or radiative shock wave.

Utilizing the initial parameters proposed by Zhang *et al.* [42], we executed simulations of the radiative blast waves problem. We established the initial gas density at $5 \times 10^{-6}$ g/cm$^3$, with the gas in a state of rest. Initial gas and radiation temperatures were set within a spherical region of radius $2 \times 10^{12}$ cm at $10^7$ K, while temperatures in the other areas were kept at $10^3$ K. The material is presumed to be an ideal gas, typified by an adiabatic index of 5/3 and a relative atomic mass of 1. The simulation incorporated one radiation group, with energy boundaries ranging from 0 to $10^6$ eV. The Planck and Rosseland mean coefficients were defined as $\kappa_{Pa} = \kappa_{Pe} = 2 \times 10^{-16}$ cm$^{-1}$ and $\chi_R = 2 \times 10^{-10}$ cm$^{-1}$, respectively. Fig. 5 illustrates the 1D profiles of gas temperature, radiation temperature, density, and radial velocity, as observed at $t = 10^6$ s within the computation region of 0 cm $< r < 10^{14}$ cm. The spatial step adopted in the simulation was $10^{11}$ cm. In the absence of analytical solutions for the radiative blast wave problem, the simulation results shown in Fig. 5 are compared with the high-resolution simulation results of Zhang *et al.* [42], which utilized a cell size of $4.9 \times 10^{10}$ cm (referenced in Fig. 14 in Ref. [42]). The comparative analysis revealed a significant consistency between our simulation results and those presented by Zhang *et al.* [42]. Concurrently, we executed 2D simulations in cylindrical coordinates, with a computation region of 0 cm $< r < 10^{14}$ cm, $-10^{14}$ cm $< z < 10^{14}$ cm. Fig. 6 presents the 2D distribution of density and gas temperature at $t = 10^6$ s, exhibiting good spherical symmetry in the simulation results.

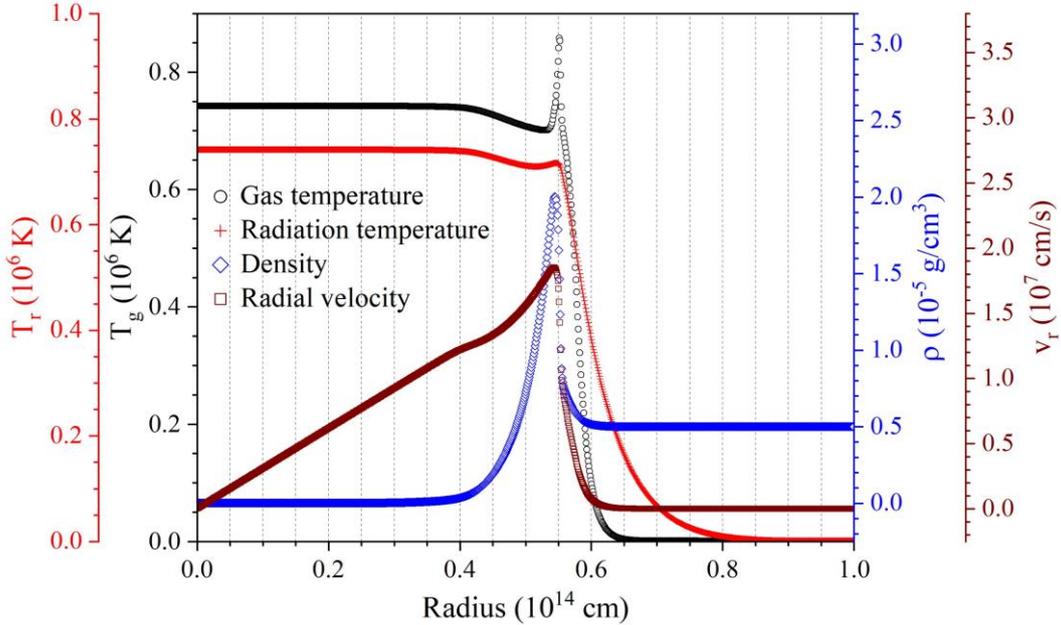

**Fig. 5.** Density (ρ, blue diamond), radial velocity ($v_r$, wine square) in the spherical radial direction, gas temperature ($T_g$, black circle), and radiation temperature ($T_r$, red cross) profiles at $10^6$ s for the radiative blast wave problem.

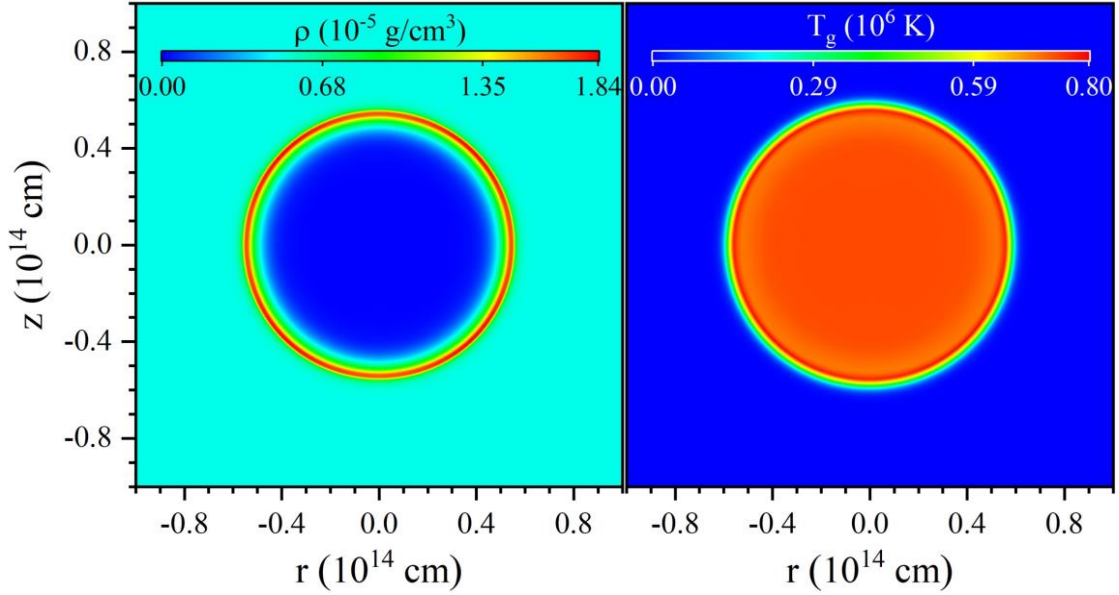

**Fig. 6.** Profiles of density (ρ, left column) and gas temperature ($T_g$, right column) of the 2D cylindrical simulation at $10^6$ s for the radiative blast wave problem.

4.3. Non-equilibrium radiative shocks

A radiative shock is a shock wave of such intensity that the radiation energy flux and pressure considerably influence its dynamics. Lowrie & Edwards have recently proposed a semi-analytic solution for radiative shocks [70]. Our simulations of the subcritical and supercritical shocks at Mach 2 and Mach 5 respectively, were juxtaposed with the solutions provided by Lowrie & Edwards for comparison.

A shock wave is classified as subcritical when the post-shock temperature exceeds the preheat temperature of the upstream material immediately preceding the shock front. A 1D simulation of the Mach 2 subcritical shock was performed within the computational domain spanning from -1000 cm to 500 cm. The initial conditions were determined based on the parameters provided by Zhang *et al.* [42]: $T_L = 100$ K, $\rho_L = 5.45887 \times 10^{-13}$ g/cm³, $u_L = 2.35435 \times 10^5$ cm/s; $T_R = 207.757$ K, $\rho_R = 1.24794 \times 10^{-12}$ g/cm³, $u_R = 1.02987 \times 10^5$ cm/s. The initial discontinuity was situated at $x = 0$, with the subscripts $L$ and $R$ indicating the left and right states of the discontinuity, respectively. The material was considered an ideal gas with an adiabatic index of 5/3 and a relative atomic mass of 1. We use 20 radiation groups with the lowest and uppermost energy boundaries at 0 and 10 eV, respectively. Initially, the radiation is assumed to be in thermal equilibrium with the gas. The group Planck and Rosseland mean coefficients were set to $\kappa_{Pa,g} = \kappa_{Pe,g} = 3.92664 \times 10^{-5}$ cm⁻¹ and $\chi_{R,g} = 0.848902$ cm⁻¹, respectively. Fig. 7 presents the density (ρ, black), gas temperature ($T_{\text{gas}}$, red), and radiation temperature ($T_r$, blue) profiles for the Mach 2 subcritical shock at $t = 0.05$ s. Here radiation temperature is defined as $(\sum_g E_g/a)^{1/4}$. The numerical simulation results are represented by circle symbols in the figure, while the solid lines denote the semi-analytic solutions by Lowrie & Edwards [70]. The numerical solutions were found to correspond closely with the semi-analytic solutions.

Supercritical shock refers to the condition where the temperature behind the shock wave matches the preheat temperature in the upstream material just before the shock front. We performed a 1D simulation of the Mach 5 supercritical shock problem in the computational domain of $-4000 \text{ cm} < x < 2000$ cm. Following Zhang *et al.* [42, 43], we adopted the following initial conditions: $T_L = 100$ K, $\rho_L = 5.45887 \times 10^{-13}$ g/cm³, $u_L = 5.88588 \times 10^5$ cm/s; $T_R = 855.72$ K, $\rho_R = 1.96405 \times 10^{-12}$ g/cm³, $u_R = 1.63592 \times 10^5$ cm/s. The setup for the radiation group remained consistent with that used for the Mach 2 simulation. Fig. 8 shows the density ($\rho$, black), gas temperature ($T_{\text{gas}}$, red), and radiation temperature ($T_r$, blue) profiles for Mach 5 supercritical shock at $t = 0.04$ s. The numerical simulation results, comprising the gas temperature with a narrow peak at $x = 0$, are highly consistent with the semi-analytic solutions.

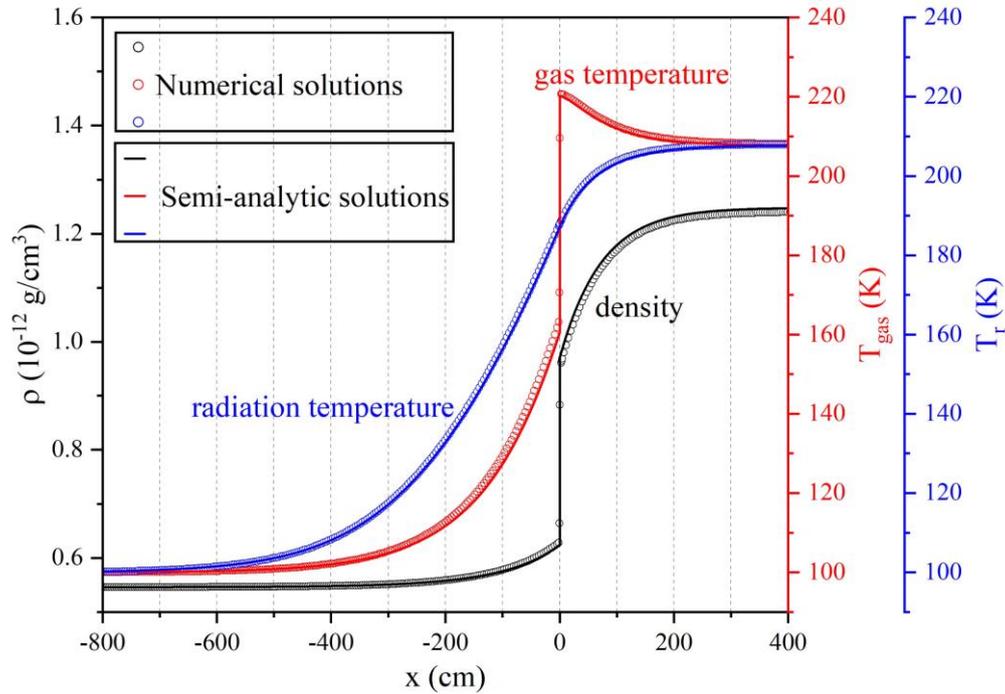

**Fig. 7.** Density (ρ, black), gas temperature ($T_{\text{gas}}$, red), and radiation temperature ($T_r$, blue) profiles at 0.05 s for Mach 2 subcritical shock. Numerical and semi-analytic solutions are shown in circle symbols and solid lines, respectively.

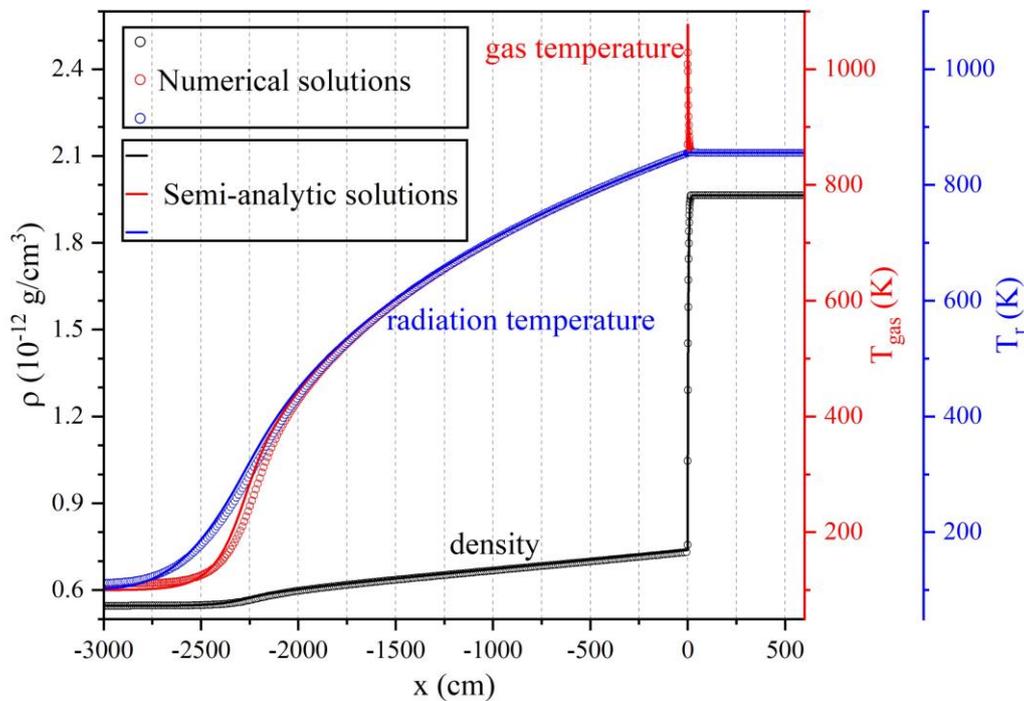

**Fig. 8.** Density (ρ, black), gas temperature ($T_g$, red), and radiation temperature ($T_r$, blue) profiles at 0.04 s for Mach 5 supercritical shock. Numerical and semi-analytic solutions are shown in circle symbols and solid lines, respectively.

## 4.4. Shock tube problem

The shock tube problem in the strong equilibrium regime, as proposed by Zhang *et al.* [42, 43], was simulated. The following initial states were established: $T_L = 1.5 \times 10^6$ K, $\rho_L = 10^{-5}$ g/cm$^3$, $u_L = 0$; $T_R = 3 \times 10^5$ K, $\rho_R = 10^{-5}$ g/cm$^3$, $u_R = 0$. Subscripts $L$ and $R$ correspond to the states at the left and right of the discontinuity, respectively, which was initially positioned at $x = 50$ cm. The material was assumed to be an ideal gas with an adiabatic index of 4/3 and a relative atomic mass of 1. The simulation was executed with 20 radiation groups, having energy boundaries ranging from 0 to $10^6$ eV. Initially, the radiation was assumed to be in thermal equilibrium with the gas. The group Planck and Rosseland mean coefficients were set to $\kappa_{Pa,g} = \kappa_{Pe,g} = 10^6$ cm$^{-1}$ and $\chi_{R,g} = 10^8$ cm$^{-1}$, respectively.

Figure 9 illustrates the gas density, velocity, total pressure, and radiation energy density profiles at $t = 10^{-6}$ s for the shock tube problem. Given the absence of an analytical solution, the 1D computational outcomes are contrasted with the radiation hydrodynamics results as well as the pure hydrodynamics findings of Zhang *et al.* (referenced in Fig. 4 in Ref. [43]), where the radiation contribution was incorporated into an equation of state for ideal gas. The comparison revealed a significant consistency between the numerical simulation results.

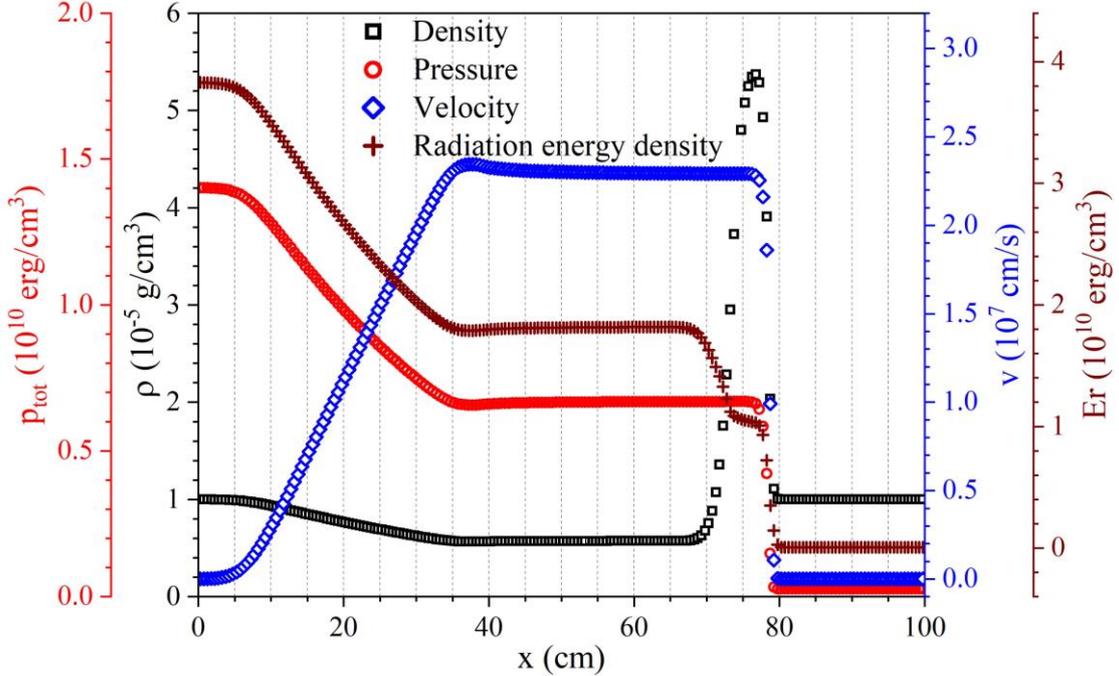

**Fig. 9.** Density ($\rho$, black square), velocity ($v$, blue diamond), total pressure ($p_{tot}$, red circle), and radiation energy density ($E_r$, wine cross) profiles at $10^{-6}$ s for the shock tube problem.

## 4.5. Linear multigroup diffusion

In this test, we focused on the linear multigroup diffusion problem as proposed by Shestakov *et al.* [71]. This problem serves as a benchmark to test the multigroup diffusion component of RHD system, particularly for verifying the numerical solution of Eqs. (23) and (24). Shestakov *et al.* [71] linearized the system under certain assumptions: absorption coefficient proportional to the inverse third power of radiation frequency, radiation emission following Wien's law instead of Planck's law, and the absence of scattering and flux-limited diffusion. These simplifications enabled them to derive exact solutions for the linearized system.

Our simulation of the first problem from Shestakov *et al.* [71] employed parameters provided in physical units by Zhang *et al.* [43]. The setup was as follows: The 1D computational domain in physical units is from 0 to $1.0285926\times 10^6$ cm with 2048 uniform cells. The time step is fixed at $5.8034112\times 10^{-8}$ s. The mass density is $1.8212111\times 10^{-5}$ g cm$^{-3}$. The initial temperature is set to 0.1 keV and 0, for $x < 0.5$ and $x > 0.5$, respectively. Here, $x$ is a dimensionless coordinate. The specific heat capacity of the matter at constant volume is $9.9968637\times 10^7$ erg K$^{-1}$ g$^{-1}$. A symmetry boundary condition is used at the lower boundary, and a Marshak boundary with no incoming flux at the upper boundary. We use 64 radiation groups with the lowest energy boundary at zero. The width of the first group is set to $5\times 10^{-2}$ eV, and the width of other groups is set to be 1.1 times the width of its immediately preceding group. The representative frequency for each group is set to $\nu_g = \sqrt{\nu_{g-1/2}\nu_{g+1/2}}$, except that $\nu_1$ is set to $0.5\nu_{3/2}$ for the first group. Here $\nu_{g-1/2}$ and $\nu_{g+1/2}$ denote the lower and upper bounds on the frequency of each group, respectively. The radiation energy density is initially set to zero everywhere. Under the first assumption, the group Planck mean coefficients for emission, and absorption, and Rosseland mean coefficient are set to be $\kappa_{Pe,g} = \kappa_{Pa,g} = \kappa_{R,g} = \kappa_g = C_\kappa \nu_g^{-3}$. The constant $C_\kappa$ is set to $4.0628337\times 10^{43}$ cm$^{-1}$ Hz$^3$. Under the second assumption, the emission term $c\kappa_{Pe,g}B_g$ in Eqs. (23) and (24) is changed to

$$c\kappa_{Pe,g}B_g = \frac{8\pi kT}{c^2}\kappa_g(\nu_g)^3\left[\exp\left(-\frac{h\nu_{g-1/2}}{kT_f}\right) - \exp\left(-\frac{h\nu_{g+1/2}}{kT_f}\right)\right]. \qquad (77)$$

where $k$ is the Boltzmann constant, the fixed temperature is $T_f = 1.16045\times 10^5$ K.

Figure 10 presents a comparison between our numerical simulation results and the exact solutions from Shestakov *et al.* [71] after 200 steps. We have represented numerical results with solid lines and the exact solutions with open circle symbols, dimensionlessized for clarity. The remarkable agreement between our results and the exact solutions validates the precision of our numerical approach.

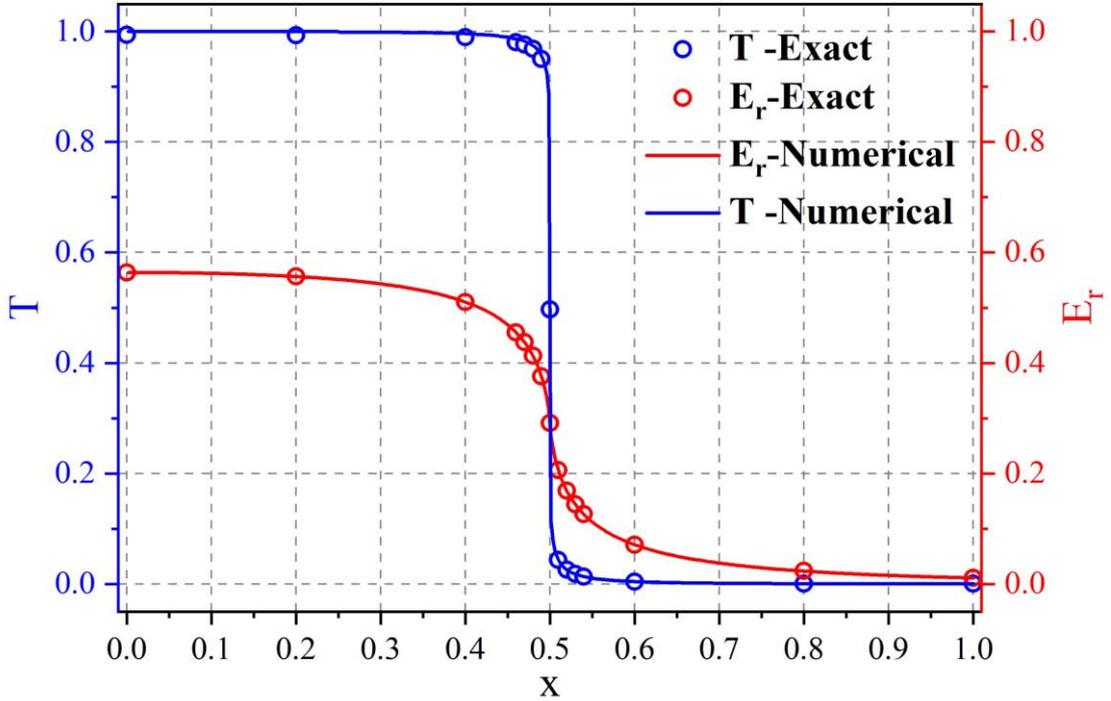

**Fig. 10.** Linear multigroup diffusion test. Numerical results are shown in solid lines, the analytic solutions are shown in open circle symbols. The red color indicates the material temperature (T) and the blue color represents the total radiant energy density (Er).

## 5. Simulations of laser-produced plasmas

In this section, we present simulation results for LPPs obtained using the RHDLPP code. The distributions of state parameters in the Al plasmas generated by ns lasers with varying powers are provided in Subsection 5.1. Subsection 5.2 focuses on the simulation of the deformation and evolution of spherical Sn microdroplets induced by both ns and ps lasers. Finally, Subsection 5.3 showcases the simulation results for imaging and spectroscopy of the Al plasma.

5.1. Laser-produced Al plasmas

The target material selected for this simulation is Al, with a thickness of 80 μm and a density of 2.7 g/cm³. The background gas used is hydrogen, with a density of $10^{-9}$ g/cm³. The initial temperature of both the target and the background gas was 300 K. Two Gaussian laser pulses were chosen for the incident laser pulses: one with an energy of 0.68 J (peak power of $4.53 \times 10^7$ W) and the other with an energy of 6.8 J (peak power of $4.53 \times 10^8$ W). Both pulses had a full width at half maximum (FWHM) of 10 ns, peaking at 15 ns, a wavelength of 1064 nm, a focal spot radius of 200 μm on the target surface, and a FWHM of 125 μm for the focal spot.

Due to the high laser intensities involved, these simulations were conducted using the RHDLPP-HTP package. The computational domain for the 2D cylindrical simulation was defined as $0 < r < 840$ μm and $0 < z < 900$ μm. For radiation modeling, the energy range from 0.1 eV to 100 keV was divided into 10 logarithmically distributed groups. The group Planck mean coefficient $\kappa_{P,g}$ and the Rosseland mean coefficient $\kappa_{R,g}$, assuming local thermodynamic equilibrium (with $\kappa_{Pe,g}$ and $\kappa_{Pa,g}$ in Eqs. (3) and (4) being equal, i.e., $\kappa_{Pe,g} = \kappa_{Pa,g} = \kappa_{P,g}$), were derived from the TOPS database [67]. Given the plasma's symmetry, the boundary condition at $r = 0$ was set as an axisymmetric condition. For all other boundaries, we applied an extrapolation with zero gradient for the plasma and a zero incoming flux boundary condition for the radiation. The heat conduction flux limiter $f$ in Eq. (7) was set to 0.1.

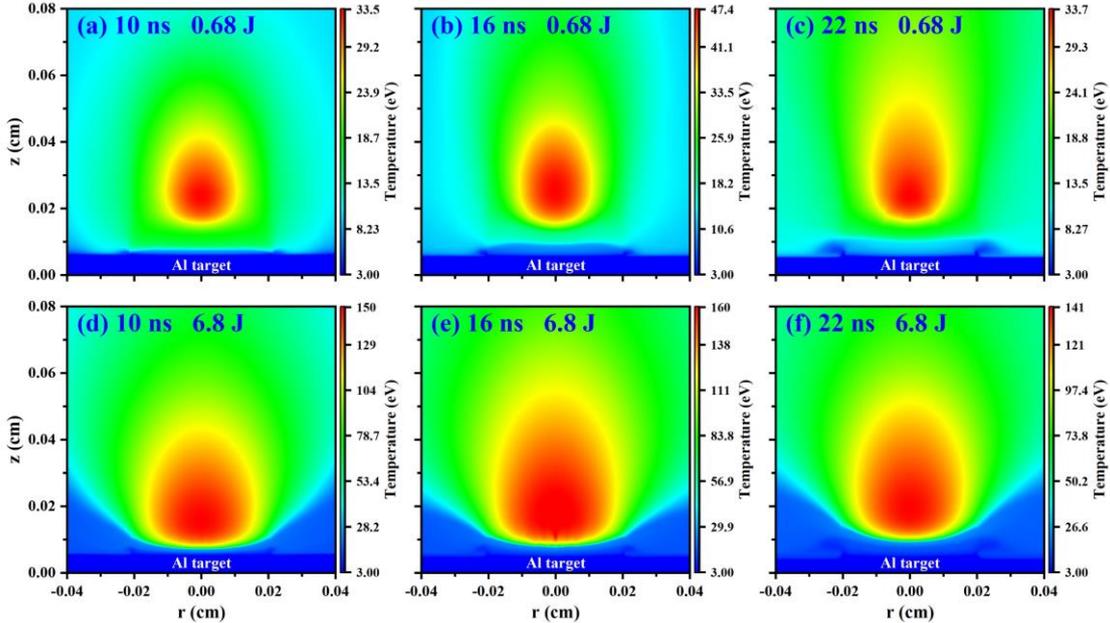

**Fig. 11.** The 2D temperature distributions of Al plasma generated by 0.68 J and 6.8 J lasers at delay times of 10 ns, 16 ns and 22 ns. The laser propagates from top to bottom in the figure.

Figure 11 illustrates the 2D temperature distributions of Al plasma produced by 0.68 J and 6.8 J laser pulses at delay times of 10 ns, 16 ns, and 22 ns. These delay times commence from the moment the laser pulse first contacts the target surface. The laser's direction of propagation is from the top to the bottom in the figure. The data reveals that the maximum temperature of the plasma generated by the 0.68 J laser increases from 33.5 eV at 10 ns to 47.4 eV at 16 ns, while the plasma from the 6.8 J laser escalates from 150 eV to 160 eV. This temperature rise is attributed to the increasing energy deposition into the plasma's interior from 10 ns to 16 ns. In the latter phase of the Gaussian temporal profile of the laser pulse (post 20 ns), the plasma temperature rapidly decreases due to factors such as diminishing laser energy, rapid plasma expansion, conversion of internal energy into kinetic energy, and thermal radiation energy loss. Consequently, at 22 ns, the maximum temperatures drop to 33.7 eV for the plasma produced by the 0.68 J laser and to 141 eV for the 6.8 J laser.

To further validate the RHDLPP code, we conducted a comparative analysis between its simulation outcomes and those obtained using the FLASH code [12], as depicted in Fig. 12. This figure presents the temperature distribution along the $z$-axis at $r = 0$, with a delay time of 14 ns. In Fig. 12(a), the laser energy is set at 0.68 J, while in Fig. 12(b), it is 6.8 J. The temperature profiles generated by the RHDLPP code are represented by red and blue solid lines, connected by circular symbols, indicating matter temperature and radiation temperature, respectively. Conversely, the FLASH code results are illustrated using black, olive, and wine solid lines, corresponding to electron temperature ($T_e$), ion temperature ($T_i$), and radiation temperature ($T_r$), respectively. The temperature distribution profiles from both codes show considerable alignment. The observed differences in detail can be attributed to several factors: The FLASH code employs a three-temperature model ($T_e \neq T_i \neq T_r$), which incorporates the energy exchange between electrons and ions, a feature not included in our model; The two codes differ in the number of radiation energy groups and the opacity parameters used; There are variations in the EOS data utilized by each code; The FLASH code exclusively uses a geometrical-optics ray-tracing method for calculating the propagation and energy deposition of the incident laser; The formulas used to calculate the frequency of electron-ion collision ($\nu_{ei}$) differ between the two codes.

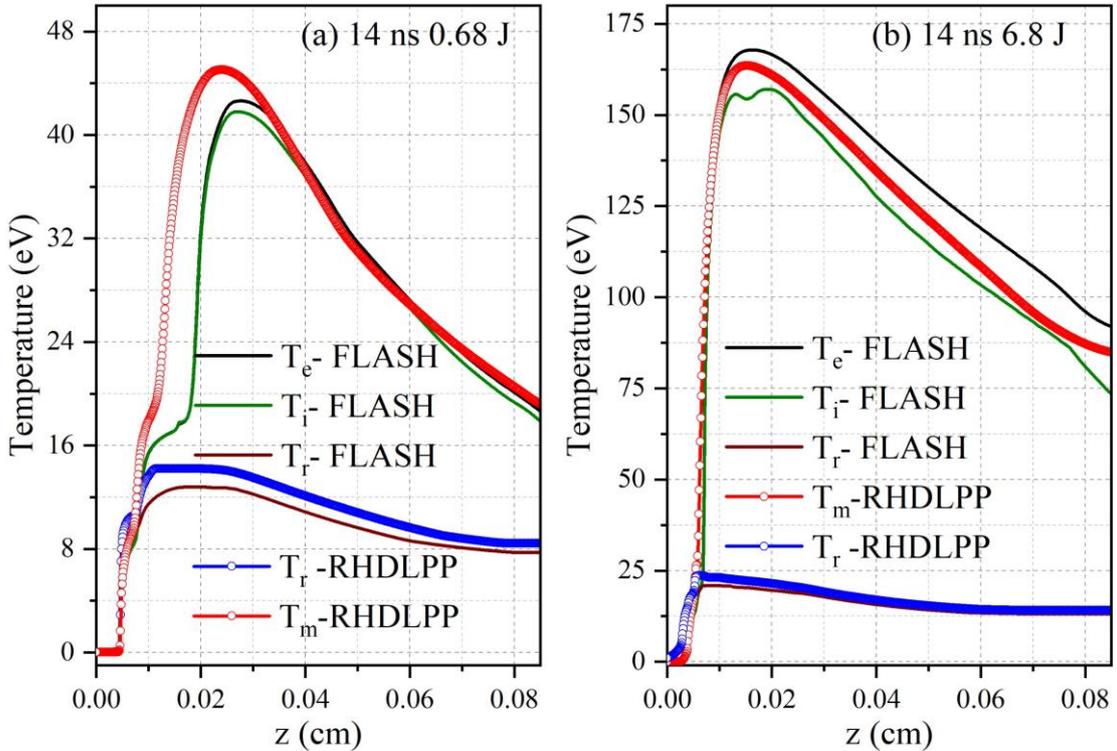

**Fig. 12.** The temperature distribution of Al plasma along the $z$-axis at $r = 0$, at a 14 ns delay, was simulated using the RHDLPP and FLASH codes. Figure (a) depicts the plasma generated by a laser energy of 0.68 J, while Figure (b) illustrates the plasma produced by a laser energy of 6.8 J.

## 5.2. Laser-induced deformation of Sn microdroplets

Current state-of-the-art nanolithography machines employ mass-limited liquid Sn microdroplets, which are effective in generating minimal debris particles [72]. These microdroplets are used to create a hot, dense plasma that emits EUV light at a wavelength of 13.5 nm. A prevalent method for generating EUV light is the dual-pulse scheme [6], involving an initial low-energy laser pulse (pre-pulse) that strikes the Sn microdroplet, deforming it into an optimal target shape. Following this, a higher-energy secondary pulse (main pulse) irradiates the shaped target, producing a dense, hot plasma. This plasma, primarily composed of $Sn^{10+}$ to $Sn^{14+}$ ions, predominantly emits EUV radiation at 13.5 nm.

Simulating both the deformation of droplets induced by the pre-pulse and the plasma generation by the main pulse is vital for enhancing the efficacy of EUV light sources. Accurate simulations offer insights into plasma parameters that are challenging to measure experimentally. They enable rapid evaluation of current source optimization strategies and enhance the understanding of the underlying physics, thereby facilitating improved experimental design. In this subsection, we will demonstrate the applicability of the RHDLPP code in the field of EUVL light sources by showing the simulation results for the pre-pulse-induced deformation of Sn droplets.

Utilizing the RHDLPP code, we conducted two simulations to model spherical Sn droplets with an initial density of 6.9 g/cm$^3$, in an environment where the background gas was hydrogen at a density of $10^{-8}$ g/cm$^3$. In both simulations, the radiation energy, ranging from 0 to 10 keV, was segmented into 20 distinct energy width groups, maintaining the heat conduction flux limiter at 0.1. The first simulation employed an unpolarized laser pulse with a 1064 nm wavelength and 6.0 mJ pulse energy. This laser had a Gaussian temporal profile with a full-width at half maximum (FWHM) pulse duration ($\tau_p$) of 10 ns, peaking at 15 ns, and a Gaussian spatial profile with a 100 μm focal spot diameter. The initial droplet radius ($R_0$) was 25 μm.

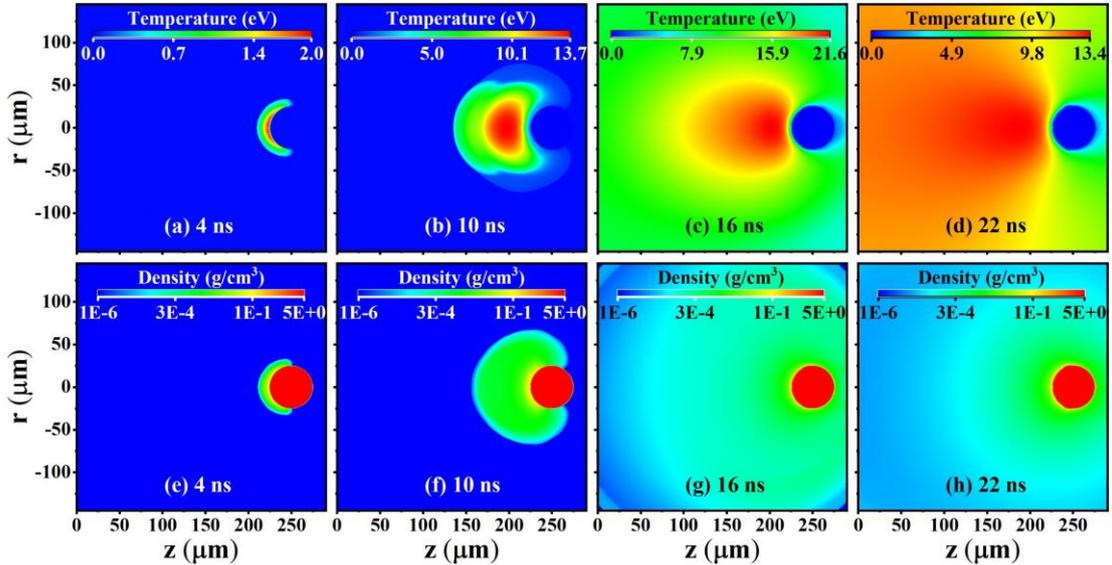

**Fig. 13.** The temperature distributions (upper row) and density distributions (lower row) at different delay times during the laser pulse impact by the first simulation. Laser pulse propagating from left to right. The center of the droplet is located at $r = 0$, $z = 250$ μm.

Figure 13 displays the temperature and density distributions during the laser pulse impact from the first simulation. In the figure, laser pulse propagating from left to right. The ns laser pulse quickly raised the temperature of the droplet surface beyond the boiling point, generating Sn vapor near the surface, as shown in Figs. 13(a) and 13(e). This vapor, produced by the laser pulse front, acted as a thin, transparent medium, permitting most of the laser beam to reach the droplet's surface. However, a fraction of the incoming laser radiation was absorbed by the vapor through electron-neutral inverse Bremsstrahlung, leading to vapor breakdown and initial plasma formation. The initial plasma then absorbed more laser energy predominantly via inverse Bremsstrahlung, significantly raising the plasma temperature and density. Given that the plasma critical density interface ($n_e = 10^{21}$ cm$^{-3}$) was close to the droplet surface, and the laser energy was mainly deposited there, the high-temperature plasma regions, as shown in Figs. 13(b), 13(c), and 13(d), were also near the surface. In the latter phase of the Gaussian temporal profile (post 20 ns), factors like reduced laser energy, rapid plasma expansion, conversion of internal energy into kinetic energy, and thermal radiation energy loss led to a swift decline in plasma temperature. Consequently, as depicted in Figs. 13(c) and 13(d), the peak plasma temperature of 21.6 eV at 16 ns dropped to 13.4 eV by 22 ns.

Figure 14 illustrates the density distribution of the deformed Sn target at various delay times following the laser off. The figure reveals that the initially spherical Sn microdroplets transform into a thin sheet-like target, approximately 10 μm thick along the z-axis, after a certain delay time, as seen in Figs. 14(f) and 14(g). This transformation aligns with results obtained from experimental measurements. The underlying physical mechanism for this deformation involves the rapid expansion of the hot plasma, generated by the ns pulse on the droplet surface, in a direction opposite to that of the laser incidence. According to the law of conservation of momentum, this expansion exerts a recoil pressure at the droplet-plasma interface. This recoil pressure induces radial expansion of the droplet and propels it along the laser propagation direction. The resultant initial radial expansion and propulsion velocity lead to significant deformation of the droplet post-laser pulse impact, indicating that the droplet's dynamic evolution progresses into the fluid dynamic response stage on the inertial time scale $\tau_i \sim R_0/U$ [6, 73, 74], where $U$ represents the droplet propulsion velocity. The value of $U$, estimated to be approximately 50 m/s from the first simulation, is consistent with experimental observations [75].

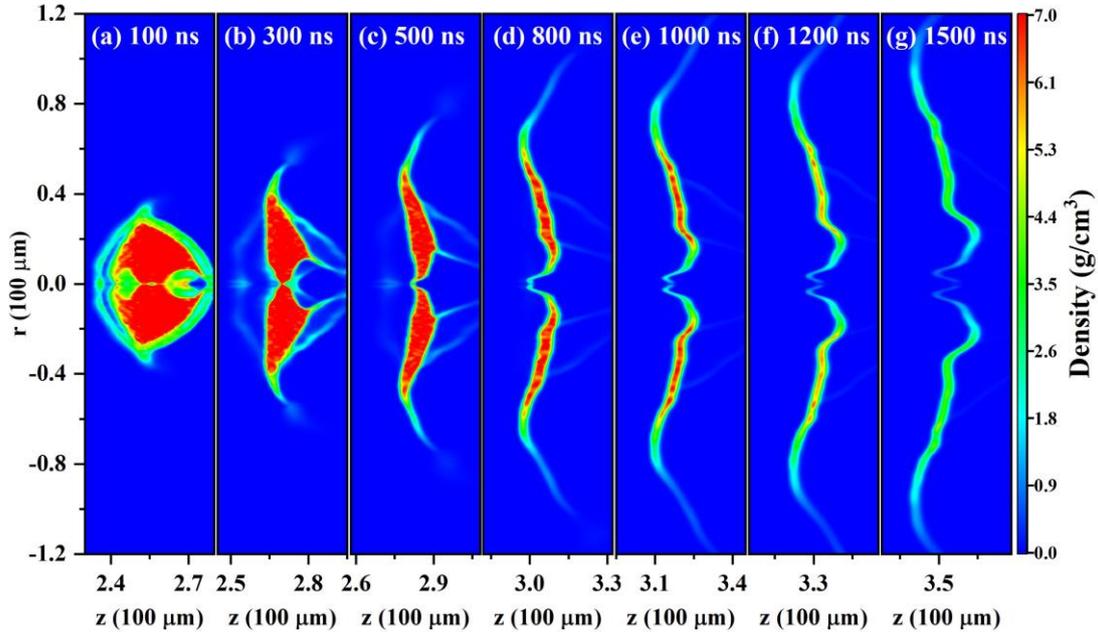

**Fig. 14.** The density distribution of the deformed Sn target at various delay times after laser off.

In the second simulation, we utilized a short pre-pulse laser with a wavelength of 1064 nm, a FWHM of 1.5 ps, and a pulse energy of 2.6 mJ. This laser exhibited a Gaussian spatial profile with a focal spot diameter of 80 μm. All other parameters remained consistent with those in the first simulation. Fig. 15 displays the density distribution of the Sn target deformed by the ps pre-pulse at delay times of 1 ns, 25 ns, and 900 ns. The center of the original spherical droplet, denoted by a solid white line in Figs. 15(a) and 15(b), is located at $r = 0$ and $z = 200$ μm.

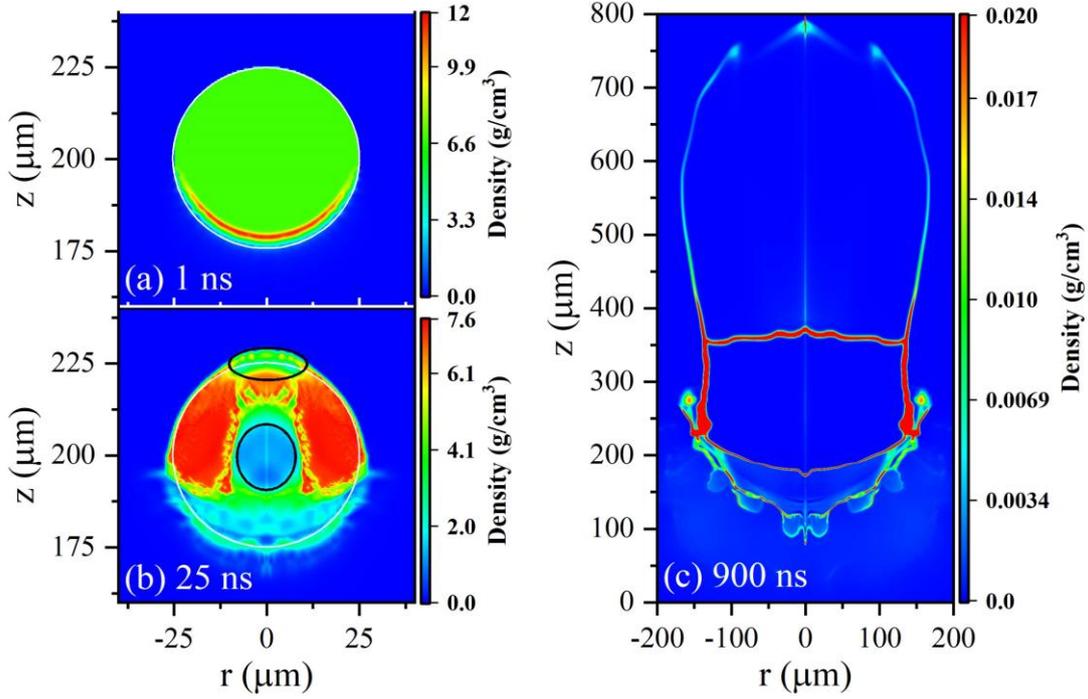

**Fig. 15.** The density distribution of the deformed Sn target induced by ps pre-pulse at delay times of 1 ns, 25 ns, and 900 ns.

Upon the ps laser pulse impacting the Sn droplet, approximately 18% of the incident laser energy was absorbed by a thin near-surface layer, about 0.1 μm thick. This rapid heating produced a hemispherical pressure wave in the laser-impacted region of the droplet, reaching up to approximately 24 GPa. Initially, as demonstrated in Fig. 15(a), the density increased within the hemispherical shell due to the pressure wave. This wave then converged towards the droplet's center, broadening spatially while the maximum pressure gradually decreased. At the convergence point in the center, the pressure and density maxima increased owing to the focusing effect. The pressure wave then passed through the center and dispersed, leading to a subsequent reduction in the maximum pressure and density values.

Furthermore, two distinct hollow structures, outlined by black solid curves, are evident in Fig. 15(b). The first hollow, at the droplet's center, resulted from a diverging velocity field created by the pressure wave, leading to the liquid's cavitation under this velocity field's pull. The second hollow was formed when the pressure wave reflected off the rear side of the droplet, rupturing the liquid again and causing spallation; a small portion of the droplet was propelled in the laser beam's direction. These processes of cavitation and spallation resulted in explosive deformation of the droplet. As depicted in Fig. 15(c), the initial spherical Sn microdroplet transformed into an acorn-like expanding shell with two inner cavities, resembling two conjunct coaxial spheroids. This shape is in good agreement with the stroboscopic shadow photography results reported in Refs. [76] and [77], and numerical simulations in Ref. [78].

In short, this subsection demonstrates the simulation capabilities of the RHDLPP code in the field of EUVL light sources by simulating the deformation of Sn microdroplets induced by ns and ps pre-pulse.

5.3. Imaging and spectral simulations of laser-produced plasmas

In this subsection, we demonstrate the capabilities of the RHDLPP code in spectral simulation by presenting simulation results for transient images and spectra of LPPs, alongside comparisons with experimental data.

Our initial simulations focused on the imaging of laser-produced Al plasmas in background air at pressures of $1\times10^5$ Pa, 2000 Pa, and $1\times10^{-5}$ Pa. For generating Al plasmas under these varying background pressures, we utilized a laser with a Gaussian-shaped pulse. The pulse characteristics included a peak intensity of $5\times10^8$ W cm$^{-2}$, a wavelength of 1064 nm, and a FWHM duration of 10 ns. The transient images were taken in the wavelength range of 200-850 nm. The simulation process initiated with the RHDLPP-LTP package determining plasma parameters like temperature and density, which were subsequently utilized in the RHDLPP-SpeIma3D module for ray-tracing and radiation intensity calculations. Extensive details on the simulation methodology are provided in Subsection 3.6.

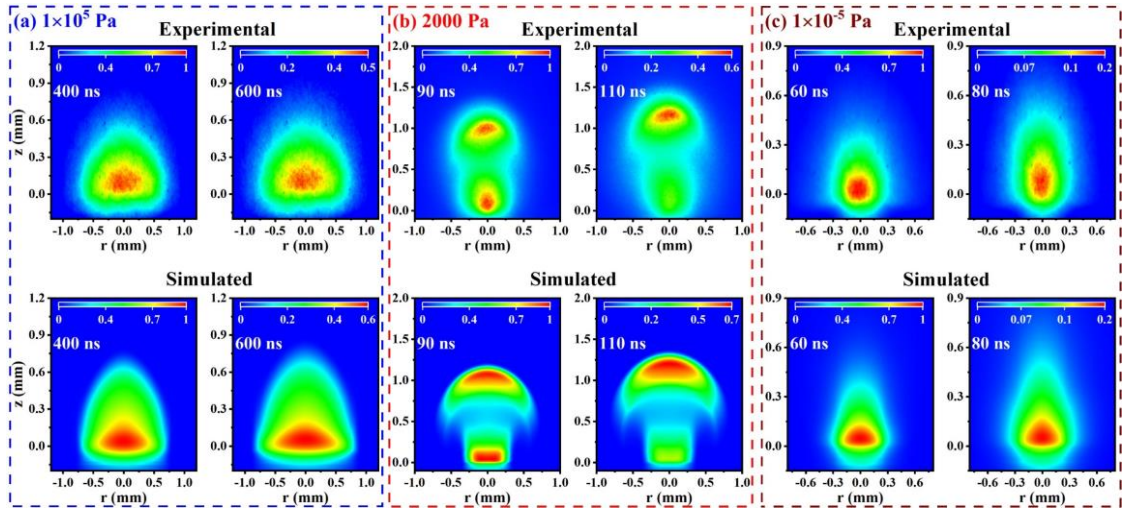

**Fig. 16.** Experimental and simulated transient images of laser-produced Al plasmas in background air with pressures of (a) $1\times10^5$ Pa, (b) 2000 Pa, and (c) $1\times10^{-5}$ Pa. The upper row displays the experimental images, while the lower row shows the corresponding simulation images.

Figure 16 showcases both experimental and simulated transient images of Al plasmas in background air at the aforementioned pressures. The figure uses blue, red, and brown dashed boxes to distinguish the different background pressures. The top row displays the experimental results, while the bottom row features the corresponding simulation outcomes. The timestamps on each image mark the acquisition time relative to the laser's impact on the target, designated as time zero. The colormaps in each image represent normalized radiation intensities within the 200-850 nm range. Detailed methodologies for these transient image measurements are documented in our previous publications [66, 79], they are not repeated in this paper for brevity.

We used three criteria to assess the agreement of the simulated images with the experimental imaging results: (1) Characteristic Reflection: The simulation results must accurately reflect the distinctive features of the experimental images. As can be seen in Fig. 16, the simulations at all three pressures reproduce the plasma profile well, especially the plume splitting at an ambient pressure of 2000 Pa. (2) Dimensional Consistency: The size of the simulated luminescent region should align with the experimental findings. This is exemplified in Fig. 16, where, for instance, at 400 ns in Fig. 16(a), the dimensions of both the experimental and simulated luminescent regions along the $z$-axis at $r=0$ are approximately 0.8 mm, at 90 ns in Fig. 16(b) they are around 1.2 mm, and at 60 ns in Fig. 16(c) they are close to 0.6 mm. This dimensional consistency is shown more clearly in the comparison in Fig. 17. (3) Intensity Distribution Consistency: The spatial distribution and temporal evolution of radiation intensity in the simulation should be consistent with experimental data, as is evident in the colormap of Fig. 16. To further clarify the intensity distribution consistency, Fig. 17 presents a comparative analysis of the experimental and simulated radiation intensities along the $z$-axis at $r = 0$, for various background pressures within the 200-850 nm wavelength range. In this figure, the experimental data are denoted by open circles, and the simulated results by solid lines. This comparison substantiates the overall agreement between the spatial and temporal distributions of the simulated intensities and the experimental findings.

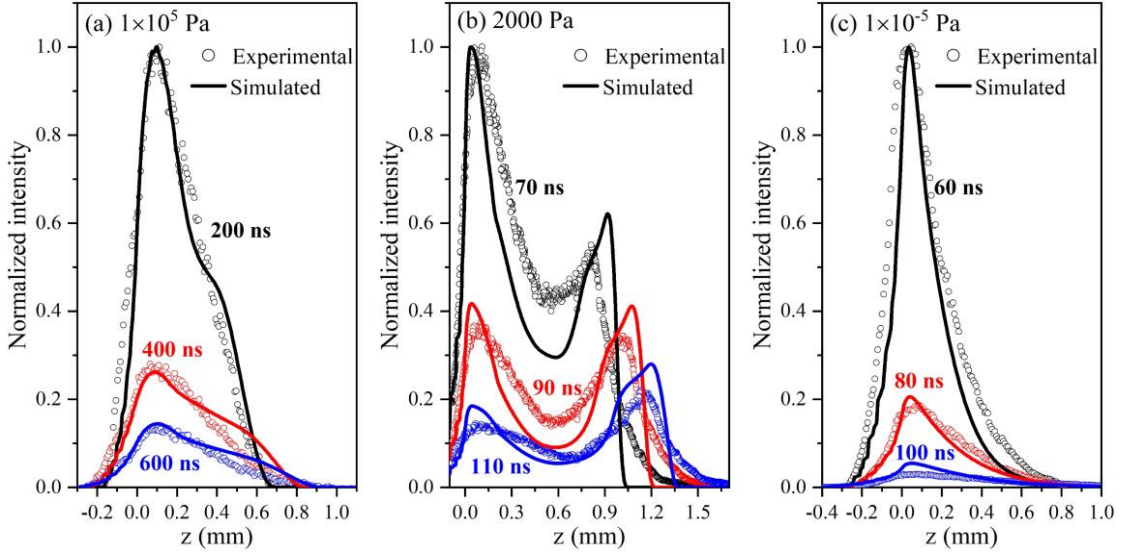

**Fig. 17.** The experimental (open circles) and simulated (solid lines) intensities along the $z$-axis at $r = 0$ for various delay times and within the 200-850 nm wavelength range. The pressures of the background air are (a) $1\times10^5$ Pa, (b) 2000 Pa, and (c) $1\times10^{-5}$ Pa.

Additionally, we conducted an EUV spectral simulation of Al plasma in the 8.5-13.5 nm range. Temperature and density distributions, crucial for these spectral simulations, were derived from plasma parameters produced by a 0.68 J laser, as calculated by the RHDLPP-HTP package (refer to Subsection 5.1). These parameters were then used in the RHDLPP-SpeIma3D module for radiation intensity calculations.

Figure 18 compares experimental and simulated Al plasma spectra in the 8.5-13.5 nm range. The experimental data are represented by black solid line, while the simulated results are depicted using red solid line. The annotations on the figure (e.g., 40 ns, 50 ns) indicate the delay times measured from the instant the laser impacts the target surface. The experimental apparatus incorporated a detector with a 50 ns gate width, enabling the capture of 50 ns time-integrated spectra. The simulation parameters were meticulously aligned with these experimental conditions. For more comprehensive experimental details, refer to our earlier publications [80]. Furthermore, the atomic structure data, essential for the EUV spectra, were computed employing the semi-relativistic Hartree-Fock method, integrated within the Cowan code [81]. In the Cowan code calculations, to refine the output, the Slater-Condon integrals $F^k$, $G^k$ and $R^k$ were reduced to 85%, while the spin parameter $\xi$ was retained. Comparative analysis revealed that the experimental spectra comprised 62 spectral lines, which originated from transitions such as 2p-3s, 3d, 4s, 4d, and 5d in Al ions ranging from $Al^{3+}$ to $Al^{6+}$.

In evaluating the congruity of the simulated spectra with experimental results, we applied three criteria, akin to our imaging approach: (1) Consistency of Spectral Features: The discrete spectral lines evident in the experimental spectra are well-replicated in the simulations, as observed in Fig. 18. (2) Consistency of Relative Intensities: At given delay times, the intensity ratios of spectral lines are consistent between the experimental and simulated spectra. For example, at 40 ns in Fig. 18, the ratio of intensities at 10.40 nm, 10.80 nm, and 10.96 nm in the experimental spectra (0.95:0.85:0.61) aligns with that of the simulated spectra. This consistency extends to other delay times. (3) Consistency of Time Evolution: The time-dependent intensity variations in the experimental spectra, such as the diminishing intensity at 10.40 nm, 10.80 nm, and 10.96 nm from 40 ns to 70 ns, are faithfully mirrored in the simulated spectra.

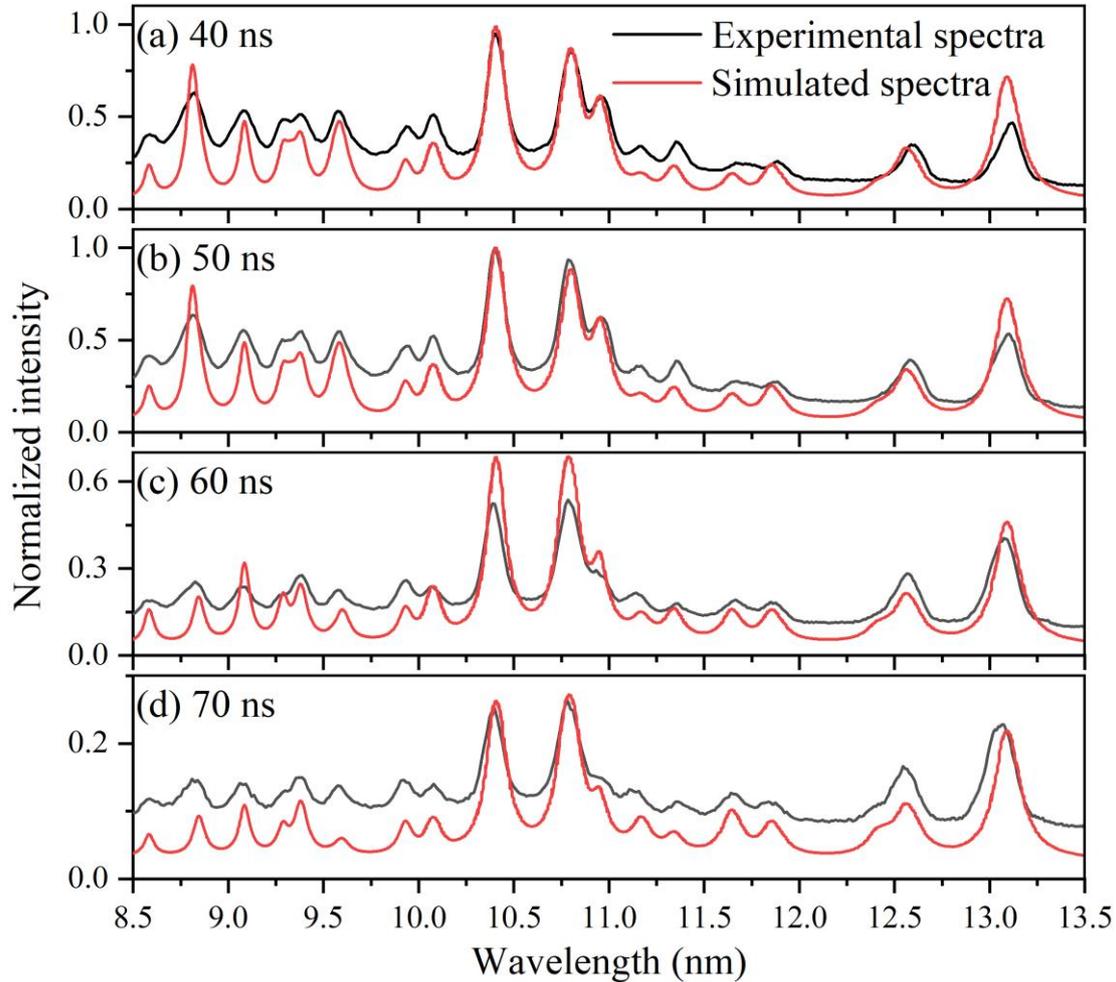

**Fig. 18.** The experimental (black solid line) and simulated (red solid line) Al plasma spectra for various delay times and within the 8.5-13.5 nm wavelength range.

Finally, we conducted simulations of EUV spectra generated by laser-produced Sn plasmas. The simulations employed a planar solid Sn target, measuring 100 μm in thickness and possessing a density of 7.3 g/cm³. The environment consisted of a hydrogen background gas with a density of $10^{-9}$ g/cm³. Initially, both the target and the background gas were maintained at a temperature of 300 K. We utilized a Gaussian Nd:YAG laser pulse, characterized by an intensity of $3\times10^{10}$ W/cm², a FWHM duration of 7 ns, a wavelength of 1.064 μm, and a focal spot radius of 275 μm on the target surface. These simulation parameters align with the experimental conditions described by Pan *et al.* [82], who conducted joint measurements of electron density, temperature, and emission spectrum in Nd:YAG laser-produced Sn plasma.

Figure 19(a) presents the 2D distributions of both temperature (top row) and electron density (bottom row) within the Sn plasma at various delay times: -2 ns, 0 ns, 2 ns, and 4 ns, with the peak laser intensity moment defined as 0 ns [82]. The original planar Sn target is positioned at $z = 100$ μm. The observed temperature and electron density distributions at 0 ns, 2 ns, and 4 ns correspond closely with the experimental data reported by Pan *et al.* [82], obtained through the Collective

Thomson Scattering (CTS) method (refer to Fig. 2(a) in Ref. [82] for experimental results). Additionally, Figs. 19(b) and (c) illustrate the 1D temperature and electron density distributions at distances of 130 μm, 200 μm, and 300 μm from the target surface, parallel to it, at a delay time of 0 ns. The hollow symbols signify our simulation outcomes, while the black, red, and blue shaded areas represent the experimental results (refer to Figs. 2(b) and 2(c) in Ref. [82]), underscoring the consistency between our simulations and the experimental findings.

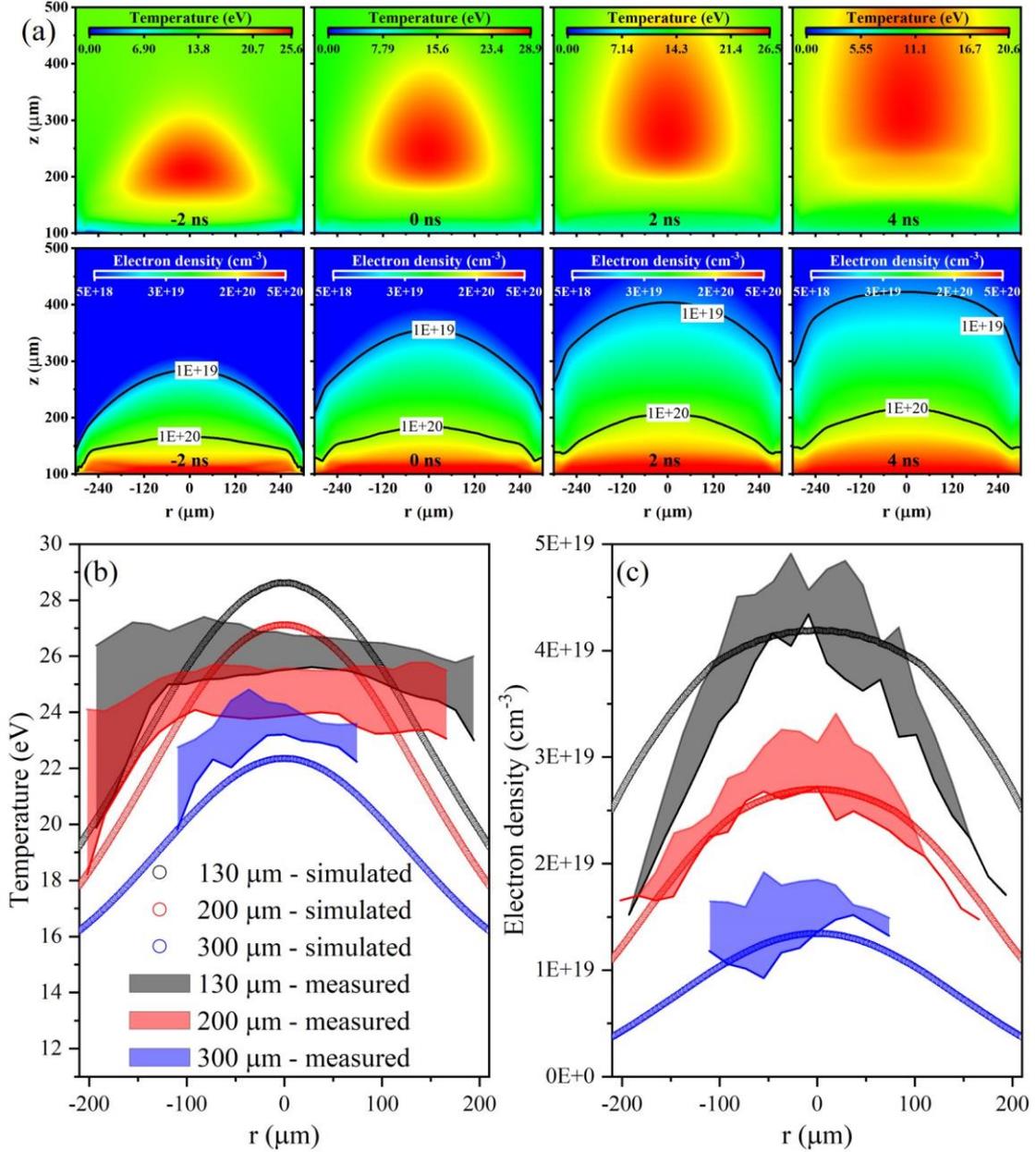

**Fig. 19.** (a) Displays the 2D simulation results for the Sn plasma. The top row depicts the temperature distribution, while the bottom row focuses on the electron density, across various delay times: -2 ns, 0 ns, 2 ns, and 4 ns. (b) & (c) Illustrate the 1D distributions. In (b), the temperature distribution, and in (c), the electron density distribution, are shown at distances of 130 μm, 200 μm, and 300 μm from the target surface. These distributions are aligned parallel to the target surface and are calculated for a delay time of 0 ns. In these figures, the hollow symbols represent the outcomes of our simulations. The experimental findings are depicted through shaded areas in black, red, and blue, as reported in reference [82].

The parameters from Fig. 19, such as temperature and electron density, were used as input conditions in the RHDLPP-SpeIma3D module to simulate EUV spectra within the 12-16.4 nm wavelength range. Figure 20 showcases the simulated and measured spectra at distances of 130 μm, 200 μm, and 300 μm from the target surface. The 2% bandwidth around the 13.5 nm wavelength, significant for nanolithography applications, is indicated by the gray-shaded area. The hollow symbols in the figures represent space-resolved experimental spectra (refer to Fig. 5 in Ref. [82]), measured by Pan *et al.* using a flat-field grazing incidence spectrometer (GIS) [82], while the solid lines depict our simulated spectra. Atomic structure parameters essential for these simulations were computed using the Cowan code [81]. Our simulation incorporated one-electron-excited states of $Sn^{7+}$ to $Sn^{14+}$ ions [83, 84] and multiply-excited states of $Sn^{11+}$ to $Sn^{14+}$ ions [85, 86], specifically within the studied wavelength range. The simulated spectra align remarkably well with the measured spectra, demonstrating the applicability of the RHDLPP code in EUVL light source simulations.

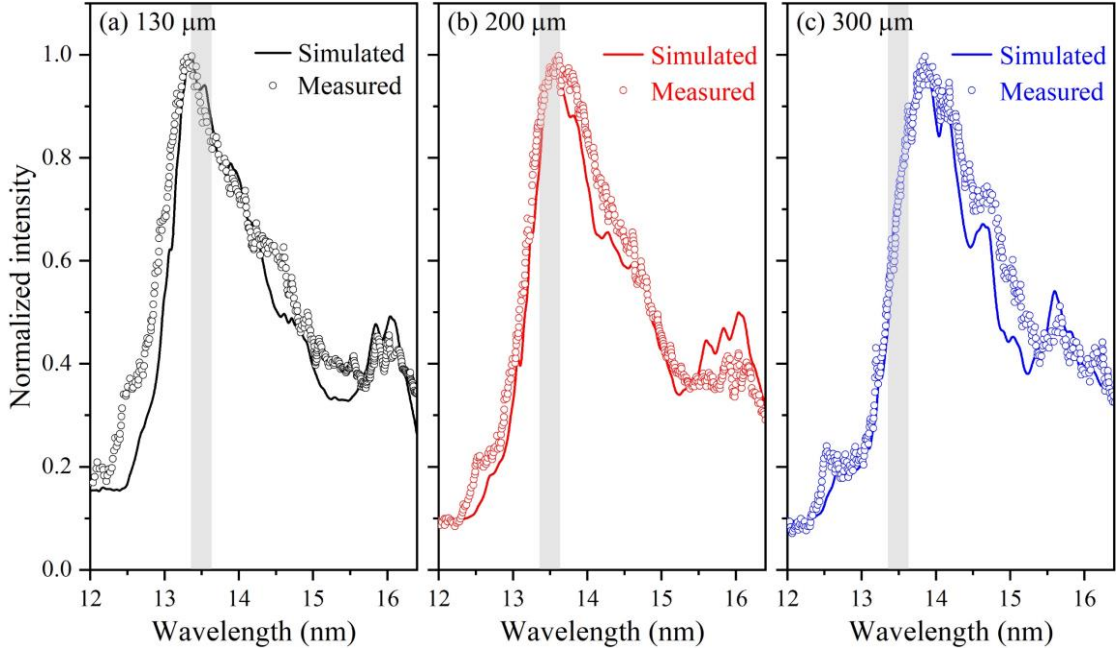

**Fig. 20.** The EUV spectra of Sn plasma at (a) 130 μm, (b) 200 μm and (c) 300 μm from the target surface. Each segment exhibits a gray-shaded area, highlighting the 2% bandwidth centered around the 13.5 nm wavelength, which holds significant importance in nanolithographic applications. The hollow symbols denote the space-resolved experimental spectra measured by Pan *et al.* using a flat-field grazing incidence spectrometer (GIS) [82]. The solid lines depict the spectra generated from our simulations.

Overall, the results presented in this subsection validate the robustness of the RHDLPP code for both imaging and spectral simulations, highlighting its potential for broader applications in fields such as LIBS, EUVL, and HEDP.

## 6. Summary

In this paper, we present the RHDLPP, a specialized radiation hydrodynamics code developed for the simulation of laser-produced plasmas. This code employs a multi-group method for solving the radiation diffusion model, using a flux-limited diffusion approximation to approximate the free-streaming limit. The RHD equations, forming a mixed hyperbolic-parabolic system with stiff source terms, are solved through an operator-split method consisting of two distinct substeps. The first substep explicitly solves the hyperbolic subsystems, integrating radiation and fluid dynamics. This is accomplished using the Strang operator split scheme, which comprises a half-step update for source terms related to radiation force, work done by the radiation field, and geometrical contributions, followed by a full step update of the hydrodynamic equations, and another half-step update for source operations. The hydrodynamic equations are resolved using a second-order TVD MUSCL-Hancock scheme. The second substep addresses the parabolic components, including stiff radiation diffusion, heat conduction, and energy exchange, through a two-stage implicit solver. This involves updating the radiation energy diffusion and energy exchange between matter and radiation groups in nested iterations, then directly updating heat conduction implicitly.

The RHDLPP code integrates modules for hydrodynamics, heat conduction, radiation transport, laser energy deposition, equations of state, charge state distribution, and spectral simulation. Laser propagation and energy deposition are modeled using a hybrid approach, combining geometrical-optics ray-tracing in sub-critical plasma regions with a 1D solution of the Helmholtz wave equation in super-critical regions. Thermodynamic properties are determined using an EOS based on either the real gas approximation or a QEOS. Charge state distribution and average ionization degree are calculated using a steady-state CR model, incorporating various atomic processes based on the screened-hydrogenic approximation. The RHDLPP-SpeIma3D, a post-processing spectral simulation module, generates images and spectra by solving the radiative transfer equation across multiple lines-of-sight through the plasma.

Extensive testing has demonstrated RHDLPP's capability to address a wide range of radiation hydrodynamics problems. We also showcase application cases for the RHDLPP code in LPP scenarios, including simulations of laser-produced Al plasmas, pre-pulse-induced target deformation in Sn microdroplets, and various imaging and spectroscopic simulations. These cases illustrate RHDLPP's effectiveness and broad applicability in areas like LIBS, EUVL sources, and HEDP.

The RHDLPP code offers two primary advantages in the simulation of LPPs. Firstly, it is bifurcated into two distinct packages: RHDLPP-LTP, designed for simulating low-temperature plasmas generated by moderate-intensity ns lasers, and RHDLPP-HTP, aimed at high-temperature, high-density plasmas produced by high-intensity laser pulses. This division enables the code to effectively simulate a wide range of LPP scenarios, applicable in various research areas. Secondly, the inclusion of the RHDLPP-SpeIma3D module, dedicated to imaging and spectral simulation, allows for direct comparisons between simulated and experimental spectral results. This feature not only facilitates the benchmarking of the code but also enables comprehensive analysis of experimental data and the generation of novel predictions.

The current version of the RHDLPP code, while robust, still has several deficiencies that need to be improved. Plans are underway to address these deficiencies, including: 1. Developing a new radiation hydrodynamics solver using Eulerian grids coupled with a block-based AMR strategy, transitioning from the current uniform grid discretization; 2. Implementing 3D laser ray tracing algorithms to overcome the constraints of 2D axisymmetric geometry in laser ray-tracing, particularly addressing issues related to non-parallel rays heating the plasma near the cylindrical axis; 3. Developing an independent code for calculating radiation opacity of plasmas in non-local thermodynamic equilibrium states using a detailed configuration accounting approach, moving away from reliance on databases such as TOPS Opacities [67] and THERMOS [54]. These advancements aim to further enhance the capabilities and accuracy of the RHDLPP code.

## Acknowledgements

This work was supported by the National Key R&D Program of China (2022YFA1602500), the National Natural Science Foundation of China (NSFC) (11904293, 12064040, 11874051, 12374384, 62335016, 12305222), the Traditional Chinese Medicine Industry Innovation Consortium Project of Gansu Province (22ZD6FA021-5), the Central Leading Local Science and Technology Development Fund Projects (23ZYQA293).